\pdfoutput=1
\documentclass{article}
\usepackage[colorlinks=true,citecolor=blue,linkcolor=blue,urlcolor=blue]{hyperref} 
\usepackage{graphicx} 
\usepackage{arxiv}
\usepackage{cite}
\renewcommand{\undertitle}{}
\makeatletter
\renewcommand{\@maketitle}{%
  \vbox{%
    \hsize\textwidth
    \linewidth\hsize
    \vskip 0.1in
    \@toptitlebar
    \centering
    {\LARGE\sc \@title\par}
    \@bottomtitlebar
    \vskip -0.15in
    \def\And{%
      \end{tabular}\hfil\linebreak[0]\hfil%
      \begin{tabular}[t]{c}\bf\rule{\z@}{24\p@}\ignorespaces%
    }
    \def\AND{%
      \end{tabular}\hfil\linebreak[4]\hfil%
      \begin{tabular}[t]{c}\bf\rule{\z@}{24\p@}\ignorespaces%
    }
    \begin{tabular}[t]{c}\bf\rule{\z@}{24\p@}\@author\end{tabular}%
  \vskip 0.15in \@minus 0.1in
  }
}
\renewenvironment{abstract}
{
  \vskip -0.05in
  \centerline
  {\large \bfseries \scshape Abstract}
  \vskip 0.05in
  \begin{quote}
  \vskip -0.05in
}
{
  \end{quote}
}
\makeatother
\usepackage{multicol}
\usepackage{multicol}
\usepackage{xcolor}
\usepackage{amssymb}
\usepackage{amsmath}
\usepackage{siunitx}

\newcommand{\Wpre}{W_{\mathrm{pre}}}
\newcommand{\Wdns}{W_{\mathrm{DNS}}^{\mathrm{3D} }}
\newcommand{\WdnsTwoD}{W_{\mathrm{DNS}}^{ \mathrm{2D} }}
\newcommand{\WdnsTwoDL}[1]{W_{\mathrm{DNS},\,L=#1}^{2D}}
\newcommand{\Wexp}{W_{\mathrm{DNS+Exp}}^{ \mathrm{2D} }}
\newcommand{\WdnsNative}{W_{\mathrm{DNS},128}^{\mathrm{native}}}
\newcommand{\Sim}{\mathcal{S}}
\newcommand{\Exp}{\mathcal{E}}
\newcommand{\affiliation}[2]{\mbox{$^{#1}$#2}}

\title{Emergent transfer of a physics foundation model from simulation to laboratory turbulence}
\author{\normalfont\mdseries
Payel Mukhopadhyay\thanks{Corresponding Email: \texttt{pm858@cam.ac.uk}}$^{1,9}$,
Stefan S. Nixon$^{1}$,
Romain Watteaux$^{2}$,
Michael McCabe$^{3,9}$,\\[0.55em]
Alberto Bietti$^{3,9}$,
Kyunghyun Cho$^{4,9}$,
Cristiana Diaconu$^{1,9}$,
Irina Espejo Morales$^{4,9}$,
David Fouhey$^{4,9}$,\\
Siavash Golkar$^{4,9}$,
Tom Hehir$^{1,9}$,
Shirley Ho$^{3,4,5,9}$,
Jake Kovalic$^{6,9}$,\\
Géraud Krawezik$^{3,9}$,
François Lanusse$^{7,9}$,
Tanya Marwah$^{9}$,
Rudy Morel$^{3,9}$\, \\
Mariel Pettee$^{8,9}$,
Helen Qu$^{3,9}$,
Jeff Shen$^{5,9}$,
Hadi Sotoudeh$^{1,9}$,\\[0.55em]
Stuart B. Dalziel$^{1}$,
Miles Cranmer$^{1,9}$\\[0.3em]
\begin{minipage}{0.95\textwidth}
\centering
\small\normalfont\mdseries
\affiliation{1}{University of Cambridge, UK;}
\affiliation{2}{CEA, DAM/DIF, 91297 Arpajon Cedex, France;}\\
\affiliation{3}{Flatiron Institute, USA;}
\affiliation{4}{New York University, USA;}\\
\affiliation{5}{Princeton University, USA;}
\affiliation{6}{Yale University, USA;}\\
\affiliation{7}{AIM, Universit\'e Paris-Saclay, Universit\'e Paris Cit\'e, CEA, CNRS, France;}\\
\affiliation{8}{University of Wisconsin--Madison, USA;}
\affiliation{9}{Polymathic AI}
\end{minipage}
}

\begin{document}

\maketitle

\begin{abstract}
Whether physics foundation models can be usefully deployed on laboratory experiments remains an open question for scientific machine learning. We test this question on the Rayleigh-Taylor instability (RTI), a ubiquitous and demanding fluid instability seen from tabletop flows to supernova explosions, in which small perturbations at a density interface grow into chaotic, multiscale mixing as a lighter fluid accelerates into a heavier one. Standard machine learning models struggle with RTI, and despite over a century of theoretical, numerical, and experimental work, it carries an unresolved discrepancy between simulation and experiment: the late-time mixing growth rate, $\alpha$, measured in most laboratory experiments ($\sim$ 0.06-0.07), is roughly three times the value from idealized direct numerical simulations (DNS, $\sim$ 0.02-0.03). The gap's origin remains debated. These properties make RTI a stringent test for a question that matters well beyond RTI: can foundation models trained only on simulations generalise to sparse, messy, and noisy laboratory settings? We finetune Walrus, a foundation model for continuum dynamics, on three or fewer DNS realizations and recover characteristic RTI physics over long rollouts. Applied zero-shot to sliding-barrier laboratory data, the finetuned model leaves the DNS-like regime and enters the observed growth band, having never seen a single experimental sample. These results provide independent, data-driven evidence that initial conditions play a crucial role in the longstanding sim-experiment gap in $\alpha$. The same model also generalises zero-shot to stable stratification, a buoyancy regime absent from training, correctly slowing mixing-layer growth. Together, our results show that foundation models can generalise well beyond their training data, predicting laboratory behavior and unseen physical regimes, opening new ways to study scientific challenges where simulations and experiments have long disagreed.

\end{abstract}

\section{Introduction}
\label{sec:intro}

Foundation models~\cite{bommasani2021foundationmodels}, neural 
networks pretrained on large, diverse datasets and subsequently 
adapted to specific tasks via finetuning, have reshaped 
natural language processing and computer vision. A key property of this paradigm, established in language and vision, is transfer: representations learned during pretraining generalise to new settings, reducing the need for large task-specific datasets. The physical sciences present a 
natural frontier for this paradigm, though whether transfer 
extends to real laboratory conditions remains much less 
established. Large simulation 
datasets~\cite{pdebench,ohana2024the,hu2026realpdebench,gupta2023towards,hao2023pinnacle,bonnet2022airfrans,toshev2023lagrangebench,2020QJRMS.146.1999H,yu2023climsim,janny2023eagle,hassan2026bubbleformer,leyli-abadi2022lips,chung2023turbulence,Li_2008,tali2024flowbenchlargescalebenchmark} and foundation models for systems governed by partial differential 
equations (PDEs) that show promising signs of generalising across 
physical 
systems~\cite{herde2024poseidon,mccabe2024multiple,subramanian2023towards,
rautela2025morph,nguyen2025physix,hao2024dpot,mccabe2025walrus,liu2024prosefd,holzschuh2025pdetransformer,cao2026vicon,rahman2024pretraining,sun2025foundationmodelpartialdifferential,morel2025disco,Serrano2026TesttimeGF,liu2026tadpoleautoencodersfoundationmodels} 
have made the pretraining-finetuning paradigm increasingly practical for physics.

Simulation is a primary tool for studying complex physical 
systems, but real laboratory experiments remain the ground 
truth, and the two do not always agree: laboratory conditions 
introduce perturbations, noise, and physical complexity that 
idealized simulations do not capture. Whether a foundation 
model finetuned on idealized simulation can generalise to 
real laboratory conditions it was never trained on, that is, 
whether it can make accurate predictions on experimental data 
without any experimental training, is a question that matters 
deeply for scientific machine learning (ML)~\cite{Channing_2026}: if 
models cannot cross this divide, their utility for real 
physical discovery is fundamentally limited. We call this 
\textit{sim-to-real} transfer. But the question goes further: 
physical systems span a vast range of regimes, many of which 
are expensive or difficult to simulate in full. Whether a 
model trained on one regime can generalise to qualitatively new physical regimes it was never shown is an open question with broad 
implications for how foundation models could be used to 
explore the physical sciences.  

Rayleigh-Taylor instability (RTI) is among the most compelling 
systems on which to explore these questions. Place a heavy fluid 
above a light one and the interface erupts into complexity: fingers 
grow, spikes plunge, bubbles rise, and what looked like equilibrium 
cascades into turbulent mixing across scales. RTI governs processes 
from inertial confinement fusion to supernova ejecta to deep-ocean 
mixing~\cite{sharp1984overview,cabotcook2006reynolds,
zhou2025instabilities}. It illustrates a challenge common across 
the physical sciences: laboratory conditions introduce perturbation 
spectra, noise, and physical complexity that idealized simulation 
does not capture, and simulation and experiment do not always agree. 
In RTI, this disagreement is well-documented and precisely 
quantified: most laboratory experiments report late-time mixing growth 
rates near $\alpha \approx 
0.07$~\cite{read1984,snider1994,ramaprabhu2004,2009PhFl...21c4103O,youngs1989_physicad,DimonteSchneider1996,dalziel1999selfsimilar,mueschke2009_pof,Glimm2013,AndrewsDalziel2010}, 
while idealized miscible direct numerical simulations (DNS) more typically yield $\alpha \approx 
0.02$-$0.03$~\cite{youngs2017_ps,dimonte2004alphagroup,cabotcook2006reynolds,youngs1994_lpb,linden1994molecular,cook2004mixingtransition,burton2011,Statsenko2013,livescu2011dns_rti,Livescu2010,Youngs2003,Youngs2013}, 
where $\alpha$ characterizes how fast the turbulent mixing layer 
grows. This gap has persisted for decades despite sustained 
theoretical, numerical, and experimental 
effort~\cite{sharp1984overview,youngs2017_ps,
zhou2017b_physrep_rtrm_ii,zhou2024_hydroinstabilities}.

Predicting RTI quantitatively at the integral and spectral level is costly and configuration-specific using 
DNS~\cite{cabotcook2006reynolds}. Widely used ML architectures struggle with RTI. Results from
The Well benchmark~\cite{ohana2024the} suggested that RTI is among the
hardest systems in the suite for popular architectures such as
FNO~\cite{li2021fourier}, TFNO~\cite{kossaifi2024multigrid}, and
U-Net~\cite{DBLP:journals/corr/RonnebergerFB15}. In our 3D RTI
setting, we observe the same qualitative picture:
standard baselines break down on RTI emulation tasks
(Appendix~\ref{app:baseline_breakdown}). RTI has long served as a 
proving ground for fluid models, from buoyancy-drag laws and 
turbulence closures to large-eddy 
simulations~\cite{sharp1984overview,zhou2017a_physrep_rtrm_i,
zhou2017b_physrep_rtrm_ii,zhou2024_hydroinstabilities}. Data-driven emulators add a further tier. Task-specific ML surrogate models have made progress on 
canonical flows such as homogeneous isotropic 
turbulence, turbulent channel flows, and related 
systems~\cite{Mohan02102020,
li2023iufno,wang2024channel,subel2021data,
guan2022stable,subel2023explaining,holzschuh2026pd,
oommen2025learning}, with encouraging generalisation 
results within the same simulation family, and on 
RTI within idealized simulation conditions~\cite{luo2024fno}. Here we take a 
fundamentally different approach, leveraging a foundation model 
pretrained on diverse multi-physics data and finetuned 
on a small number of RTI simulations, and study 
generalisation to real laboratory conditions and 
qualitatively new physical regimes.

RTI is, in many ways, a worst case: chaotic, 
multiscale, carrying a persistent sim-experiment discrepancy, and 
hard to model with popular ML architectures. Together, these properties make it a compelling testbed for a question that matters well beyond RTI itself: whether foundation models finetuned on idealized simulation can
generalise to settings they were never trained on, from real laboratory conditions to qualitatively new
physical regimes. Crucially, for RTI, the simulation-experiment gap in $\alpha$ provides 
a precise, quantitative signal of whether transfer from simulation 
to real laboratory conditions has succeeded, and success here would 
suggest that this paradigm can carry over to the broader class of 
physical systems that share the same challenges: laboratory 
conditions that resist faithful simulation and experimental data 
that is expensive to collect.

\begin{figure}[htb!]
    \centering
    \includegraphics[width=\linewidth]{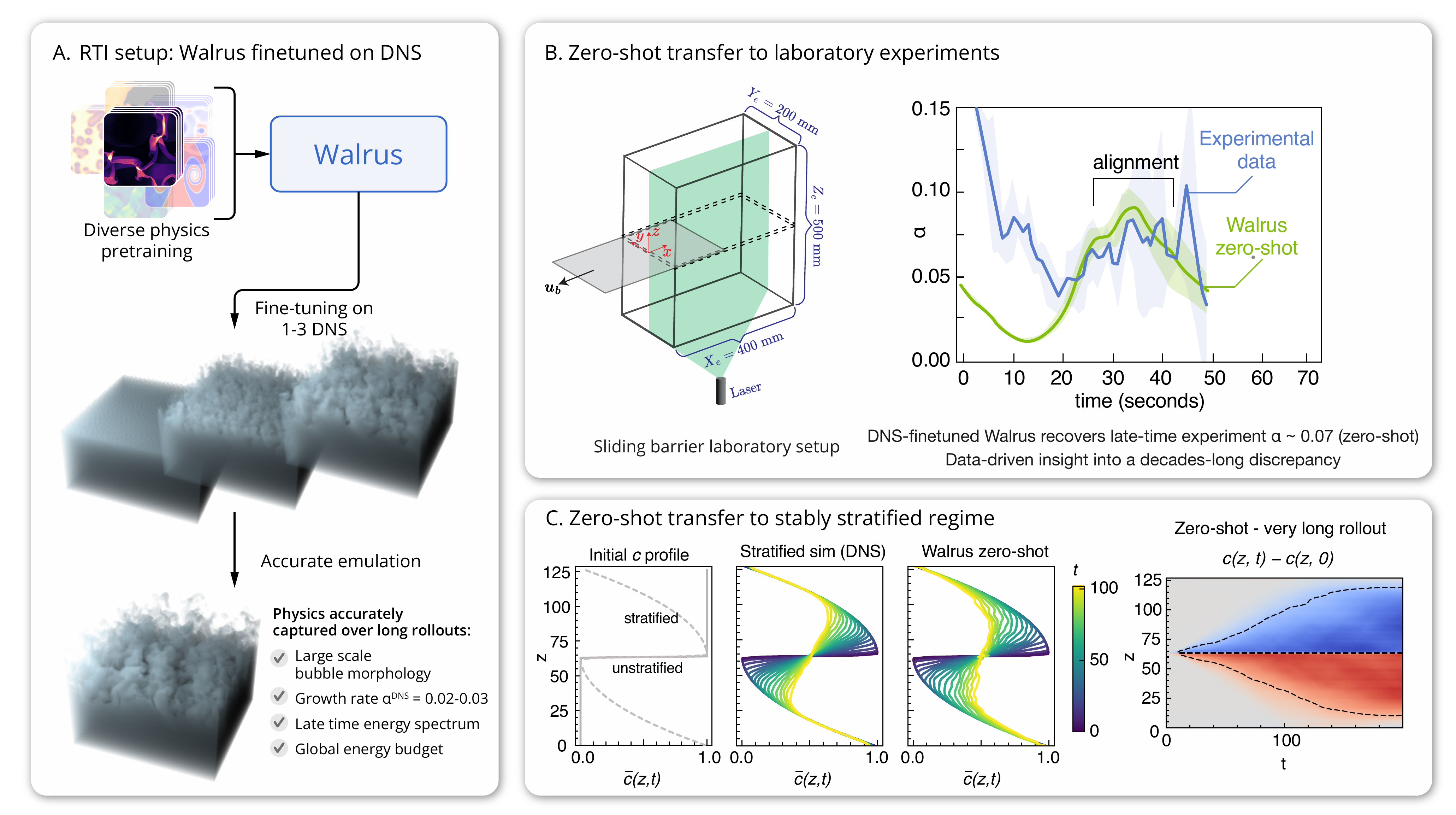}
    \caption{\textbf{Overview.}
\textbf{(A)} Walrus is pretrained on a broad set of 
continuum-dynamics simulations and then finetuned on a 
small number of RTI direct numerical simulations (DNS) 
in the Boussinesq regime (approximately incompressible, 
with small density differences driving buoyancy). From 
a total of 5 DNS realizations, we use 3 for finetuning, 
1 for validation (used to monitor training progress and 
select the best model), and 1 as a statistically 
independent held-out test realization, evaluated by 
feeding each predicted state back as input to generate 
the next one (known as autoregressive rollout), mimicking how a traditional simulation 
steps forward in time. We additionally probe how few 
DNS realizations are needed by finetuning with as few 
as 1 or 2 realizations (thereby the 1-3 range denoted in Fig.~\ref{fig:overview}A). The model 
correctly captures standard RTI physics: large-scale 
bubble morphology, mixing-layer growth rate 
$\alpha^{\mathrm{DNS}} \approx 0.02$, energy spectra, 
and the global energy budget. See 
Sec.~\ref{sec:results_emulation} for details.
\textbf{(B)} Zero-shot transfer to sliding-barrier 
laboratory experiments, 
meaning the model is applied directly to experimental 
data without any experimental training of any kind, 
neither during pretraining nor finetuning. Laboratory setup is shown here and described further in Appendix \ref{app:experimental_methods}. The model 
is finetuned only on idealized DNS and supplied with 
the first few frames of a real experiment as input, 
from which it predicts the subsequent evolution. Two 
lines are shown: the held-out experimental data (blue) 
and the Walrus rollout supplied with experimental 
frames as input (green). Early-time agreement is not 
expected at zero-shot, since the sliding-barrier transient is absent 
from the DNS training data. The decisive comparison is 
the late-time, approximately self-similar regime, 
marked as the \textit{alignment} region in the figure, 
where the DNS-experiment discrepancy in $\alpha$ is 
defined: most laboratory experiments report 
$\alpha \approx 0.06$-$0.07$, roughly 
three times the idealized DNS value 
$\alpha^{\mathrm{DNS}} \approx 0.02$. In the alignment 
region, the green and blue lines align, with the 
zero-shot rollout entering the experimentally observed 
growth band near $\alpha \approx 
0.07$, without the model ever having seen 
experimental data of any kind, neither during 
pretraining nor finetuning. See Sec.~\ref{sec:ood} 
for details.
\textbf{(C)} Zero-shot transfer to stably stratified 
RTI, meaning the model is applied directly to a 
physically distinct regime it was never trained on. 
A model finetuned only on unstratified DNS is supplied 
with stably stratified initial conditions absent from 
training. The model correctly slows and confines the 
mixing layer, matching the qualitative response of the 
stratified DNS reference, suggesting the model has 
encoded the physics of buoyancy-driven flow. See 
Sec.~\ref{subsec:zero-shot-stably-stratified} for details.
Together, these results show that foundation models 
finetuned on idealized simulation can generalise well 
beyond their training conditions, predicting both real 
laboratory behavior and qualitatively new physical 
regimes. These results suggest the 
pretraining-finetuning paradigm as a promising
path for the physical sciences~\cite{Channing_2026}. }
\label{fig:overview}
\end{figure}

The origin of this $\alpha$ discrepancy has been the subject of community debate for several 
decades, with multiple candidate explanations 
proposed~\cite{brown2018exploring,youngs2017_ps,doi:10.1073/pnas.032568799}. These explanations largely fall into three categories. The first and leading candidate is initial conditions. Laboratory flows carry 
large-scale perturbation structure set by the experimental 
apparatus that is notoriously difficult to model with DNS~\cite{DALZIEL1993127,dalziel1999selfsimilar,ramaprabhu_dimonte_andrews2005_initialperturbations,youngs2017_ps}, and both 
experiments initialised with simulation-like initial conditions and 
simulations initialised with experiment-like conditions show 
reduced 
discrepancy~\cite{2009PhFl...21c4103O,dalziel1999selfsimilar,
ramaprabhu_dimonte_andrews2005_initialperturbations,youngs2017_ps}. The second is the Schmidt number, $Sc = \nu/D$, where $\nu$ is the kinematic 
viscosity and $D$ is the mass diffusivity: laboratory flows have 
$Sc \sim O(10^3)$ while many simulations use $Sc \sim O(1)$, so 
molecular diffusion smooths density differences far more in the 
numerics than in the laboratory~\cite{brown2018exploring}. A 
third proposed explanation is numerical interfacial diffusion: standard simulation codes artificially smear the density jump at the fluid interface, 
reducing the effective buoyancy force and hence $\alpha$; non-diffusive front-tracking codes have been found to obtain higher growth rates~\cite{doi:10.1073/pnas.032568799}. Numerical studies that incorporate experimentally representative 
initial conditions into simulation recover growth rates closer to 
laboratory 
values~\cite{dalziel1999selfsimilar,
ramaprabhu_dimonte_andrews2005_initialperturbations,
mueschke2009_pof,youngs2017_ps}, but each such approach requires 
a bespoke numerical setup.

We ask whether a foundation model finetuned only on 
idealized DNS, supplied with real experimental 
conditions at inference time, can predict the late-time 
laboratory growth regime without any experimental 
training data, that is, in a zero-shot fashion. As we will show, the transfer results 
provide evidence that initial conditions indeed play an important role in driving the $\alpha$ discrepancy~\cite{dalziel1999selfsimilar,DALZIEL1993127,
ramaprabhu_dimonte_andrews2005_initialperturbations}, offering an independent, data-driven and complementary perspective on a debate that has resisted resolution through simulation alone.

To study these questions, we work with 
\textit{Walrus}~\cite{mccabe2025walrus}, a foundation 
model for continuum dynamics pretrained on broad 
simulation data with RTI excluded from pretraining (Fig.~\ref{fig:overview}A). We 
denote this pretrained model $\Wpre$ and verify that it 
fails to capture the flow structure or any canonical 
diagnostics when applied directly to RTI data; RTI 
physics must be explicitly learned through finetuning. 
We finetune $\Wpre$ on incompressible DNS in the 
Boussinesq regime~\cite{dimonte2004alphagroup} (Atwood 
number $A_t < 0.1$) (Fig.~\ref{fig:overview}A). From a total of five 3D DNS 
realizations, we use three for finetuning, one for 
validation (used to monitor training progress and 
select the best model), and one as a held-out test 
case, that is, a statistically independent realization 
never seen during training. The finetuned model is evaluated by 
initializing it from the first few frames of the 
held-out DNS and rolling out autoregressively, feeding 
each predicted state back as input to generate the 
next one, mimicking how a traditional simulation steps 
forward in time. We evaluate against RTI physics diagnostics: flow morphology, $\alpha(t)$, 
kinetic-energy spectra, and the global energy 
budget~\cite{sharp1984overview,cabotcook2006reynolds}. 
We find that the finetuned model faithfully emulates 
RTI physics across all these diagnostics (Fig.~\ref{fig:overview}A). We further 
find that finetuning on a single DNS realization already 
yields excellent spectral emulation, showing that the 
paradigm works even under extreme data constraints 
(Fig.~\ref{fig:overview}A, 
Sec.~\ref{sec:results_emulation}). 

Next we ask one of the central \textit{sim-to-real} questions of the study: can the representations learned from finetuning on idealized DNS generalise to noisy, real-world experimental data that are notoriously difficult to model with simulations? For this, we 
finetune $\Wpre$ on 2D idealized DNS slices, matching the planar measurements available from 
the laboratory, which represent an exceptionally sparse 
dataset of only six samples 
(Fig.~\ref{fig:overview}B; Appendix~\ref{app:experimental_methods}). The finetuned model is then supplied with the first few frames of sliding-barrier RTI 
experiments~\cite{DALZIEL1993127,dalziel1999selfsimilar} 
as input, where the perturbation spectrum is set by 
residual barrier motion, diffusion, phase coherence, 
and apparatus-specific noise that idealized DNS does 
not capture, and rolls out autoregressively from there without any experimental training data. The model's 
predictions enter the late-time growth band observed 
in the laboratory near $\alpha \approx 0.07$, 
without a single experimental training sample. That 
this happens from experimental initial frames alone provides direct evidence on the important role of initial conditions in driving the discrepancy~\cite{2005PhPl...12e6301D,
ramaprabhu_dimonte_andrews2005_initialperturbations,
youngs2017_ps,zhou2017a_physrep_rtrm_i}, independent 
of prior numerical approaches. We focus on the 
late-time, approximately self-similar regime because 
that is where the gap in $\alpha$ is defined; early-time 
agreement is not expected at zero-shot, since the sliding-barrier 
transient is absent from the DNS finetuning data (see 
Sec.~\ref{sec:ood}; Fig.~\ref{fig:overview}B). With only two experimental 
samples, the model can be further adapted through a 
lightweight second finetuning stage to rapidly learn 
the dominant features of the experimental initial 
conditions, qualitatively capturing the complex large-scale perturbation structure set by the barrier release that, as described above, has resisted modeling with DNS~\cite{dalziel1999selfsimilar,
ramaprabhu_dimonte_andrews2005_initialperturbations,
mueschke2009_pof,youngs2017_ps}.

Beyond the sim-to-real question, the successful emulation results (Fig.~\ref{fig:overview}A) raise a further question: if the model has learned the underlying physics of RTI, can it generalise to a physically distinct regime 
that demands a robust representation of the underlying physics of buoyancy, given RTI is fundamentally a buoyancy driven flow? We test this by initializing the finetuned model (which only saw unstratified, idealized DNS) with the first few input frames of a stably stratified flow, where a stable density gradient acts as a 
restoring force that suppresses and eventually arrests 
the mixing layer rather than driving it, a fundamentally 
different buoyancy regime entirely absent from training. We find that the model correctly captures the suppression and confinement of the mixing layer characteristic of the stratified regime, suggesting 
the learned representation has encoded the physics of 
buoyancy-driven flow at a level general enough to extend 
beyond the training distribution (Fig.~\ref{fig:overview}C). Together, these results show 
that foundation models finetuned on idealized simulation 
can generalise well beyond their training conditions, 
predicting both real laboratory behavior and new physical regimes. 

\textbf{Contributions}. \textbf{(1)} We demonstrate 
that the pretraining-finetuning paradigm can achieve 
sim-to-real transfer and generalise to qualitatively 
new physical regimes in a demanding, data-limited 
setting, suggesting it as a promising approach for 
scientific ML problems where laboratory data is scarce 
and idealized simulation falls 
short. We 
show this on RTI, which as we argue above is among the 
most stringent and physically significant systems on 
which to test this capability. \textbf{(2)} 
\textit{Walrus}, an ML foundation model, pretrained on broad simulation data 
and finetuned only on idealized DNS, transfers 
zero-shot to sliding-barrier laboratory data, 
entering the experimentally observed late-time growth 
regime near $\alpha \approx 0.06$-$0.07$ without a 
single experimental training sample. Without the ability to generalise to real laboratory 
conditions, physics foundation models are fundamentally 
limited in their utility for real physical discovery. 
Our results show that this generalisation is achievable 
on RTI, one of the most challenging systems in fluid 
dynamics. 
As a direct consequence, the zero-shot transfer provides independent, and
purely data-driven evidence that initial conditions play a crucial role in the longstanding sim-experiment discrepancy in
\(\alpha\), consistent with the inital conditions explanation~\cite{2005PhPl...12e6301D,
ramaprabhu_dimonte_andrews2005_initialperturbations,
youngs2017_ps,zhou2017a_physrep_rtrm_i}. Our approach therefore offers a new
perspective on a debate that has resisted resolution
through simulation alone. \textbf{(3)} With only two 
experimental samples, the same model can be further 
adapted to qualitatively capture the dominant 
structures of the experimental initial conditions in 
a purely data-driven way, a problem that has proved 
notoriously difficult to reproduce in DNS and that 
prior approaches required bespoke numerical setups to 
address~\cite{dalziel1999selfsimilar,
ramaprabhu_dimonte_andrews2005_initialperturbations,
mueschke2009_pof,youngs2017_ps}. \textbf{(4)} 
Finetuning on three or fewer DNS realizations produces 
a physically faithful emulator of RTI turbulence, 
recovering the integral, spectral, and energetic 
diagnostics that define the problem, showing that the 
paradigm works under severely data-limited settings. 
The same model transfers zero-shot to stably stratified 
RTI, a fundamentally different buoyancy regime absent 
from all training data, suggesting the model has 
encoded robust representations of the underlying 
physics of buoyancy, general enough to extend to qualitatively new 
physical regimes. Physically faithful emulation, sim-to-real transfer and zero-shot generalisation to new physical regimes is one of the central promises of foundation models for scientific ML~\cite{Channing_2026}, and our results suggest this promise is within reach.

These results speak to a question broader than RTI itself. Many 
physical systems present the same challenges: laboratory 
conditions that resist faithful numerical capture, 
experimental data that is expensive to collect, and a 
persistent gap between simulation and experiment. RTI is, in many ways, a worst case of this class. That foundation models 
succeed here suggests that data-driven AI approaches can open 
new routes to problems in the physical sciences that have resisted resolution for decades.

Throughout this paper, \textit{experiment} refers to a real-world 
laboratory experiment, not a numerical study. We sometimes use the abbreviated notation ``exp'' to denote experiment.

\section{Background on Rayleigh-Taylor Instabilities}
\label{sec:background}

RTI evolution is characterized through its late-time, 
approximately self-similar growth regime, in which the 
macroscopic mixing-layer thickness $h(t)$ follows the 
classical scaling $h(t) \sim \alpha A_t g t^2$,
where $A_t = \frac{\rho_1 - \rho_2}{\rho_1 + \rho_2}$ 
is the Atwood number, $\rho_1$ and $\rho_2$ are the 
initial densities of the upper and lower layers 
respectively, $g$ is the acceleration, and $\alpha$ is 
the dimensionless late-time growth coefficient that 
characterizes how fast the mixing layer 
grows~\cite{fermi_vonneumann1953_taylorinstability,
youngs1984_numericalsimulation_rti,sharp1984overview}.
We measure $h(t)$ using the standard concentration-based 
definition~\cite{cabotcook2006reynolds}
\begin{equation}
h(t) \;=\; 6 \int_0^{L_z} \bar{c}(z,t)\,
\big[1-\bar{c}(z,t)\big]\,dz,
\label{eq:mixing_width}
\end{equation}
where
\begin{equation}
\bar{c}(z,t) \;=\; \frac{1}{H_xH_y}\int\!\!\int 
c(x,y,z,t)\,dx\,dy
\label{eq:c_zavg}
\end{equation}
is the horizontally averaged concentration profile 
($H_x, H_y$ are the domain lengths in the $x$- and 
$y$-directions; $z$- direction is perpendicular to the interface). This definition weights the partially 
mixed region most strongly and tracks the bulk growth 
of the layer rather than the detailed geometry of any 
single interface contour. Concentration is defined from density by
\begin{equation}
c = \frac{\rho-\rho_{2}}{\rho_{1}-\rho_{2}},
\label{eq:conc_definition}
\end{equation}
Ristorcelli and 
Clark~\cite{ristorcelli_clark2004_jfm_rtti} derived a numerically robust formulation for $\alpha$, 
obtaining the ordinary differential equation
\begin{equation}
    \dot{h}^2 = 8\,\alpha\,A_t\,g\,h,
    \label{eq:alpha_implicit}
\end{equation}
from horizontally averaged equations under a 
self-similarity assumption, where $\dot{h} = dh/dt$ is 
the instantaneous growth rate of the mixing layer. Note that a prefactor of 8 is used here, rather than the conventional 4, because $h$ represents the full mixing width rather than the independent half-widths of the bubbles or spikes.
Rearranging, $\alpha$ is defined explicitly as
\begin{equation}
\alpha(t) = \frac{\dot{h}^2}{8\,A_t\,g\,h},
\label{eq:alpha}
\end{equation}
Integrating eq.~\ref{eq:alpha_implicit} 
recovers the familiar quadratic growth law, while in 
practice a virtual time origin $t_0$ is often introduced 
to account for early-time transients and isolate the 
late-time growth rate $\alpha$. At sufficiently high Reynolds number, both 
high-fidelity 
simulations~\cite{cabotcook2006reynolds} and physical 
experiments~\cite{dalziel1999selfsimilar} exhibit a 
classical Kolmogorov $k^{-5/3}$ ($k$ denotes the frequency) kinetic-energy cascade 
over the inertial range. Recovering both the growth rate 
and the spectral signature is therefore a stringent test 
for any learned representation of RTI.

The measured value of $\alpha$ depends sensitively on how 
the instability is seeded. Dalziel et 
al.~\cite{DALZIEL1993127,dalziel1999selfsimilar} and Ramaprabhu et 
al.~\cite{ramaprabhu_dimonte_andrews2005_initialperturbations} 
showed that enriching the initial conditions with low-$k$ 
(long-wavelength) content increases the growth coefficient. 
When the interface is seeded only with short-wavelength 
perturbations, as is done in idealized DNS, the flow must build larger structures 
through nonlinear mode coupling and bubble merger, which 
acts as a bottleneck and pushes the growth rate toward a 
lower bound. When long-wavelength modes are present from 
the start, they grow rapidly and independently, driving 
$\alpha$ upward \cite{youngs2017_ps}. Growth is ultimately limited by horizontal 
confinement: once the dominant bubble scale approaches the 
transverse domain size, the self-similar cascade is 
truncated. 

Laboratory flows are often thought to carry 
more low-$k$ structure than the idealized short-wavelength 
initial conditions typically used in DNS (Fig.~\ref{fig:ic_comparison}). 
This sensitivity to the initial conditions frames a longstanding 
benchmark in the RTI literature: most laboratory experiments 
report late-time growth rates near $\alpha \approx 
0.06$-$0.07$~\cite{read1984,snider1994,ramaprabhu2004,
2009PhFl...21c4103O,youngs1989_physicad,DimonteSchneider1996,
dalziel1999selfsimilar,mueschke2009_pof,Glimm2013,
AndrewsDalziel2010}, while idealized miscible simulations 
with short-wavelength initial perturbations more typically 
lie in the range $\alpha \approx 
0.02$-$0.03$~\cite{DALZIEL1993127,dalziel1999selfsimilar,ramaprabhu2004,2009PhFl...21c4103O,
youngs2017_ps}. As described in the introduction
(Sec.~\ref{sec:intro}), this is the initial-conditions
explanation for the $\alpha$ discrepancy.

\section{Finetuning \textit{Walrus} on RTI physics}
\label{sec:FT_walrus_on_RTI}

\textit{Walrus} is a cross-domain foundation model for continuum
dynamics~\cite{mccabe2025walrus}, pretrained on a diverse suite of 
two- and three-dimensional simulation scenarios drawn from the 
\texttt{Well} collection~\cite{ohana2024the}. In this work, we 
pretrain Walrus without RTI included in the pretraining set (see 
Appendix~\ref{app:pretraining_wpre}), so that RTI physics must 
be learned through finetuning. We denote this pretrained checkpoint 
$\Wpre$. Every result in the paper follows the same pipeline: 
starting from $\Wpre$, we finetune on a small dataset of RTI 
simulations or experiments, and evaluate by rolling out 
autoregressively from held-out test samples which the model never saw during finetuning. The specific 
data and finetuned model variants are introduced in each section 
when first used.

\paragraph*{Finetuning.}
Walrus is trained as a next-step delta-prediction model on a sequence
of discrete timesteps, so \(t \in \mathbb{Z}\) labels successive 
states rather than continuous physical time. The fluid state at 
timestep \(t\) is
\[
\mathbf{q}_t = \big(c_t,\mathbf{v}_t\big),
\]
where \(c_t\) is the concentration field (Eq. \ref{eq:conc_definition}) and
\(\mathbf{v}_t=(v_{x,t},v_{y,t},v_{z,t})\) is the velocity field,
nondimensionalized by the free-fall velocity scale \(\sqrt{A_t g H}\), 
where \(A_t\) is the Atwood number, \(g\) is the acceleration, and 
\(H\) is the domain length. 

We write the generic Walrus 
time-stepping model as $\mathcal{W}_\theta$, where $\theta$ denotes 
the model parameters. Training examples are constructed from \(L\) 
consecutive input states, also called the input context length in ML terminology.  Given 
\(\mathbf{q}_{t-L+1},\mathbf{q}_{t-L+2},\dots, \mathbf{q}_t\), 
Walrus predicts the increment from timestep \(t\) to timestep \(t+1\):
\begin{equation}
\Delta \hat{\mathbf{q}}_{t+1}
=
\mathcal{W}_\theta\!\left(
\mathbf{q}_{t-L+1}, \mathbf{q}_{t-L+2}, \dots, \mathbf{q}_{t}
\right),
\end{equation}
from which the next state is reconstructed as
\begin{equation}
\hat{\mathbf{q}}_{t+1}
=
\mathbf{q}_t + \Delta \hat{\mathbf{q}}_{t+1}.
\end{equation}
The model parameters \(\theta\) are optimized by minimizing the mean 
absolute error (MAE) loss~\cite{mccabe2025walrus}. A hat denotes a 
model prediction rather than a ground-truth field. Further details 
of the finetuning procedure are in 
Appendix~\ref{app:finetuning_3d}.

\paragraph*{Inference.}
After finetuning, the model is rolled out autoregressively: the first 
\(L\) frames are provided as input, the model predicts the next 
state, and that prediction is fed back together with the most recent 
\(L-1\) states to generate the following one. Generating a 
100-step 3D rollout takes about 150 seconds on a single H100 GPU.
 

\section{RTI emulation results}
\label{sec:results_emulation}

Before studying the sim-to-real transfer and the transfer to qualitatively new physical regimes, we first show that $\Wpre$ can be finetuned to faithfully learn RTI physics in a severely data-limited regime. For a scientific machine learning model, strong transfer is
only meaningful if the source-domain physics
has first been learned reliably.

\subsection{Setup.}
\label{subsec:emulation_setup}

We denote the five three-dimensional DNS realizations by
$\Sim^{3D}=\{\Sim_1,\Sim_2,\Sim_3,\Sim_4,\Sim_5\}$.
The set $\Sim^{3D}$ consists of five $256^3$ direct numerical 
simulations generated with the TurMix3D 
code~\cite{watteaux2011_detection} using idealized initial conditions. 
Following Thevenin et al.~\cite{THEVENIN2025134947}, the initial 
interface perturbations are characterized by an annular spectrum in 
Fourier space. The perturbation is parameterized by the perturbation 
Reynolds number ($Re$), bandwidth ($B$), and steepness ($S$), which 
is determined by the mean wavenumber ($k_0$), spectral bandwidth 
($\Delta k$) and root-mean-squared amplitude ($\eta_0$) and take 
values $\{Re, B, S\} = \{7,0.3,0.1\}$ in line with other comparable 
simulations~\cite{dimonte2004alphagroup}, please see 
Appendix~\ref{app:Initial_Conditons}. The discrete Fourier modes 
within this annulus are populated with randomly sampled phases, and 
conjugate symmetry is then enforced so that the inverse Fourier 
transform yields a real-valued interface perturbation. These initial 
conditions are distinct from the radially log-normal profiles in 
Fourier space used to initialize the examples of RTI in the 
\texttt{Well}~\cite{ohana2024the}. Each realization in $\Sim^{3D}$ 
uses a different random initialization of the perturbation spectrum.

We use three samples, $\Sim_1,\Sim_2,\Sim_3$ for the training set, $\Sim_4$ for 
validation, and $\Sim_5$ as the statistically independent held-out 
test case, meaning a realization that did not form part of the 
training or validation set. Starting from $\Wpre$, we finetune on the three training samples to obtain $\Wdns$, the 3D finetuned 
model used throughout this section. The input context length used for finetuning is set at $L=3$. We evaluate the predictions of $\Wdns$ against standard RTI physics diagnostics: qualitative large scale bubble morphology (Sec.~\ref{sec:results:morphology}), energy 
spectra (Sec.~\ref{sec:results:spectra}), the mixing-layer growth 
rate coefficient, $\alpha(t)$ (Sec.~\ref{sec:results:alpha}), and the global 
energy budget (Sec.~\ref{sec:results:energy}). To assess sample 
efficiency, we additionally finetune from $\Wpre$ using only one or 
two 3D DNS realizations instead of all three, probing whether the 
pretrained prior can be specialised under severe data constraints 
(Sec.~\ref{sec:sample_efficiency}).

\subsection{Qualitative evolution and morphology}
\label{sec:results:morphology}

In laboratory RTI experiments, the measured flow field 
is an approximation of the true underlying flow, 
recorded at a finite resolution set by the diagnostic 
instrument; the full resolution of the underlying flow 
is never directly accessible. To reflect this, our goal 
is for the model to learn RTI physics at a coarser 
effective resolution, with the underlying dynamics 
sampled from a higher-fidelity flow. We therefore 
downsample the $256^3$ DNS realizations in $\Sim^{3D}$ 
to $128^3$ by block averaging over non-overlapping 
$2^3$ cells, and these downsampled fields define the 
training, validation, and test data for $\Wdns$. This 
non-overlapping approach mirrors physical grid 
coarsening in finite-volume simulations: it ensures local 
conservation by avoiding the unphysical ``double 
counting'' of mass and momentum that overlapping 
windows would introduce, and is a standard methodology 
for mapping high-resolution fields to coarse grids in 
data-driven fluid dynamics~\cite{fukami2019super}. This choice also aligns with
our broader aim of learning the physically meaningful coarse-grained
RTI dynamics, rather than fine-scale details specific to a particular
numerical discretization or native-grid realization. We compare the
rollout of \(\Wdns\) on the held-out test case \(\Sim_5\) against the
corresponding reference DNS in Fig.~\ref{fig:rti_4panel}. Results from a model trained
natively at \(128^3\) are provided for completeness in
Appendix~\ref{app:128x3_native}. 

\begin{figure}[htb!]
    \centering
    \includegraphics[width=1.0\linewidth]{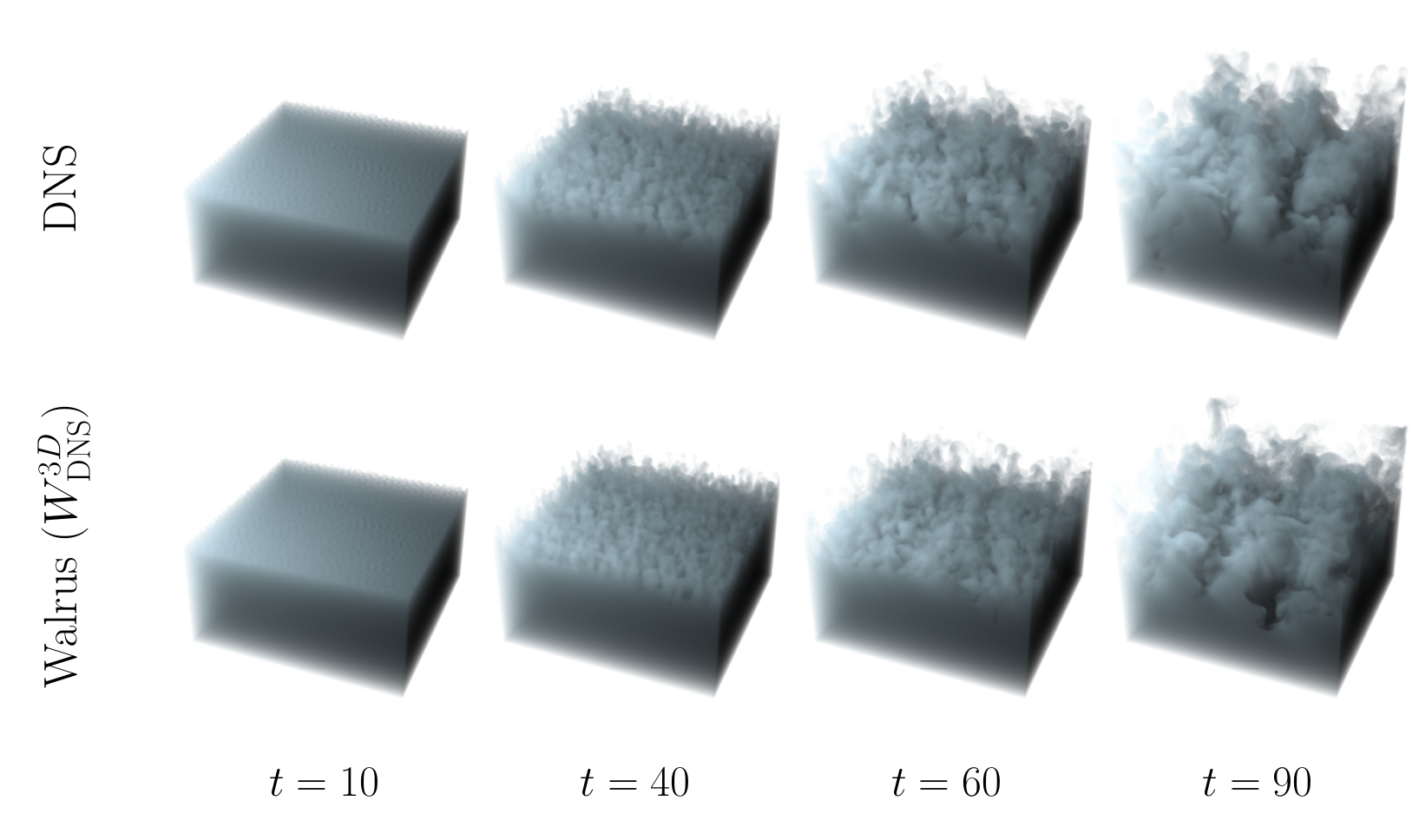}
    \caption{\textbf{Time evolution of the instability.}
Four representative times from the rollout of $\Wdns$ on the held-out
test case $\Sim_5$. Walrus tracks the growth of the mixed layer and
preserves dominant plume structures through the nonlinear and turbulent
stages.}
\label{fig:rti_4panel}
\end{figure}

As shown in Figure~\ref{fig:rti_4panel}, $\Wdns$ predicts the 
emergence of bubble morphology, the transition to the nonlinear 
regime, and the development of a well-formed mixed layer across 
representative times. As the instability grows, $\Wdns$ remains 
closely aligned with DNS in the thickening of the mixing layer and 
the persistence of the dominant plume structures through the onset of turbulence. We also compare against three widely used surrogate
architectures from The Well benchmark~\cite{ohana2024the}: FNO~\cite{li2021fourier}, 
TFNO~\cite{kossaifi2024multigrid}, and ConvNeXt-UNet~\cite{liu2022convnet}. 
For the RTI setting studied here, all three fail qualitatively under 
rollout, developing unphysical artifacts that overwhelm the true 
instability dynamics (see Appendix~\ref{app:baseline_breakdown}).

\subsection{Turbulent cascade}
\label{sec:results:spectra}

\begin{figure}[htb!]
    \centering
    \includegraphics[width=\linewidth]{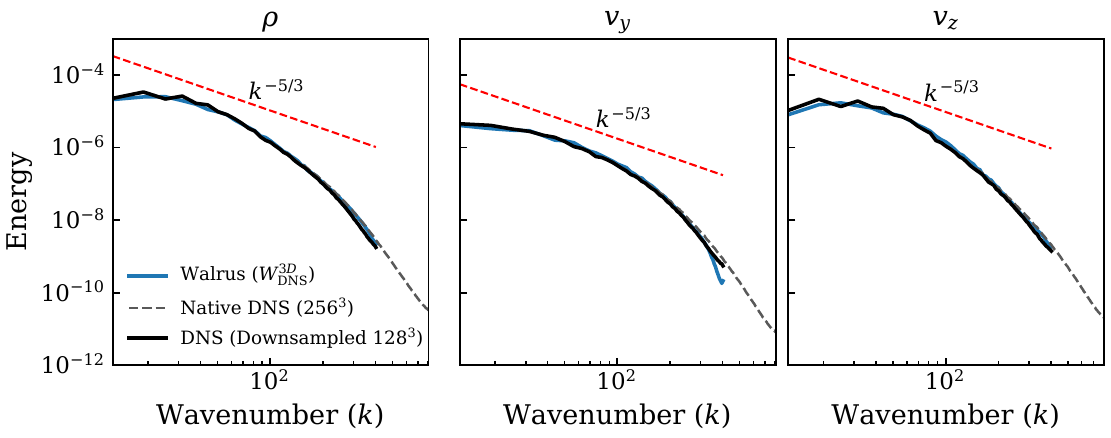}
    \caption{\textbf{Energy spectra.}
    Spectra of the density and velocity components at $t=70$, 
in the self-similar regime, comparing DNS and $\Wdns$ 
(evaluated at $128^3$; original DNS at $256^3$). Walrus 
matches the spectral shape across the inertial range, 
supporting faithful reproduction of the kinetic-energy 
distribution across scales. The $z$-direction is 
perpendicular to the interface; $x$ and $y$ are the 
two horizontal directions parallel to the interface, 
which are interchangeable for our purposes, thus only $y$- axis is shown.}
    \label{fig:cascade_zavg}
\end{figure}

The downsampled $128^3$ representation introduced above is used as the
primary reference for the spectral diagnostics that follow. As
Fig.~\ref{fig:cascade_zavg} shows, the downsampled DNS (solid black)
closely tracks the spectral shape of the native $256^3$ simulation
(dashed grey) across the inertial range, with appreciable departures
appearing only near the highest resolved modes, where the coarser-grid
cutoff begins to matter. This confirms that the
$256^3 \rightarrow 128^3$ construction retains the large- and
intermediate-scale dynamics most relevant to RTI mixing at the working resolution of the emulation task.

We next ask whether $\Wdns$ captures the distribution of kinetic
energy across scales in the self-similar regime ($t = 70$). We compute
spectra from the predicted density and velocity fields and compare them
with the downsampled $128^3$ DNS at matched time; $v_x$ is omitted
because it is statistically equivalent to $v_y$ in this geometry.

$\Wdns$ closely follows the downsampled DNS across the full resolved 
band for all three fields $(\rho,\, v_y,\, v_z)$, reproducing both 
the spectral shape and the wavenumber-by-wavenumber amplitude from 
the energy-containing scales through the inertial range 
(Fig.~\ref{fig:cascade_zavg}). The $k^{-5/3}$ Kolmogorov reference 
slope~\cite{kolmogorov1941local} is shown for orientation. Noticeable 
departures are confined to the highest wavenumbers near the grid 
cutoff, where the Walrus spectrum rolls off somewhat faster than the 
DNS reference for $v_y$.

Long-horizon autoregressive rollouts are difficult for learned PDE 
surrogates: errors compound over time and predictions often diverge 
from ground-truth simulations after only $t = 10$-$20$ rollout 
steps~\cite{ohana2024the,lippe2023pderefiner,mccabe2024multiple} 
(see also Appendix~\ref{app:baseline_breakdown}). Sustained spectral 
fidelity deep into the rollout is therefore a stringent measure of 
whether the adapted model preserves the turbulent cascade. 

\subsection{Mixing growth rate coefficient: $\alpha(t)$}
\label{sec:results:alpha}

\begin{figure}[htb!]
    \centering
    \includegraphics[width=\linewidth]{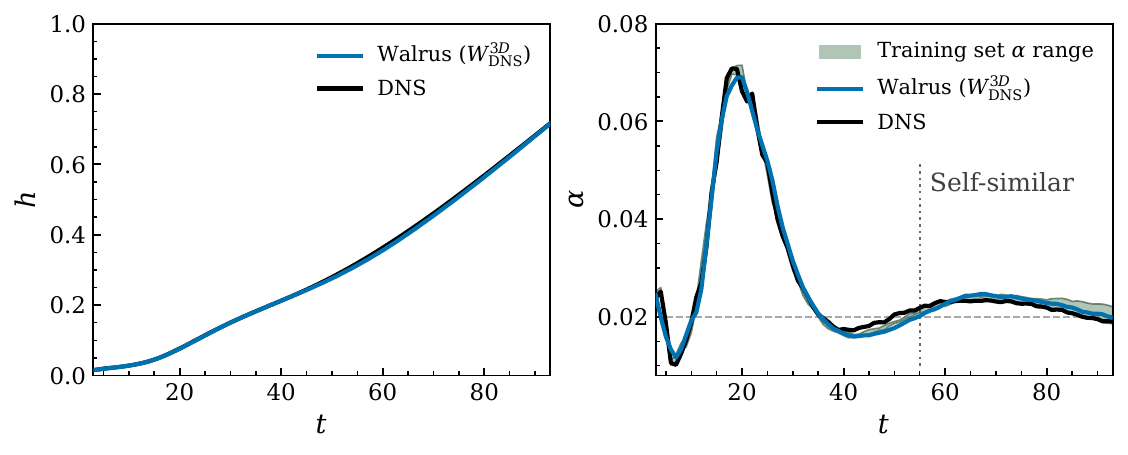}
    \caption{\textbf{Mixing growth rate.}
    Left: concentration-derived mixing width $h(t)$ computed from
    Eq.~\ref{eq:mixing_width}. Right: corresponding growth coefficient
    $\alpha(t)$ computed from $h(t)$ and its time derivative
    (Eq.~\ref{eq:alpha}) for $\Wdns$ and the held-out DNS test case
    $\Sim_5$. $\Wdns$ captures both the integrated mixing-layer growth
    and the late-time self-similar growth regime of the DNS reference.}
    \label{fig:alpha_curve}
\end{figure}

The central quantity of late-time RTI growth is the growth rate coefficient $\alpha$, as defined in Eq. \ref{eq:alpha}. In Figure. \ref{fig:alpha_curve}, we plot both the mixing layer thickness $h(t)$ (Eq. \ref{eq:mixing_width}) and $\alpha(t)$.

Figure~\ref{fig:alpha_curve} shows that $\Wdns$ closely tracks the DNS
evolution of $h(t)$ over the rollout window, indicating that the
integrated growth of the mixing layer is captured well. The more demanding quantity to model correctly is $\alpha(t) \propto \dot{h}^2 / h$
(Eq.~\ref{eq:alpha}). Because it depends on the time derivative of
$h(t)$, and in particular on its square, $\alpha(t)$ is much more
sensitive than $h(t)$ itself to small errors and fluctuations in the
predicted mixing width. Even so, $\Wdns$ follows the DNS $\alpha(t)$ curve closely
through the early evolution and into the late-time window marked in
Fig.~\ref{fig:alpha_curve}, where the flow is approximately
self-similar and where $\alpha$ is most naturally interpreted and
compared across RTI studies.

In that regime, both the held-out DNS test case $\Sim_5$ and $\Wdns$
settle near $\alpha \approx 0.02$, with the Walrus prediction
remaining very close to the DNS reference and within the spread of the
training realizations shown for comparison. The late-time difference
between $\Wdns$ and the held-out test case $\Sim_5$ is smaller than
the spread in $\alpha$ across the training realizations
$\Sim_1,\Sim_2,\Sim_3$. These values also lie within the range
commonly reported for idealized miscible RTI with short-wavelength
initial perturbations, $\alpha \approx 0.02$-$0.03$
~\cite{ramaprabhu2004,2009PhFl...21c4103O,youngs2017_ps}. This agreement is notable because the model is not trained to optimize
either $h(t)$ or $\alpha(t)$ directly. Recovering them requires the model to preserve the integrated consequences of entrainment and bulk
mixing across the full layer, together with the global flow
morphology.

For completeness, we note that Walrus predictions exhibit a small amount of run-to-run stochastic variation due to the 
patch-jittering procedure introduced 
in~\cite{mccabe2025walrus}; the conclusions of the 
emulation analysis are robust to this stochasticity, 
as shown in Appendix~\ref{app:robust_stochastic}.

\subsection{Energetics: kinetic energy and released potential energy}
\label{sec:results:energy}

\begin{figure}[htb!]
    \centering
    \includegraphics[width=\linewidth]{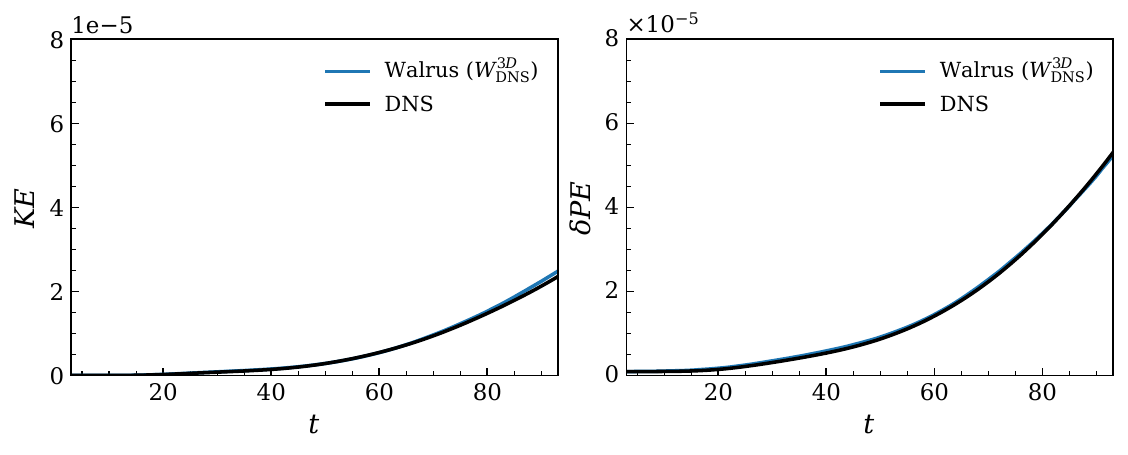}
    \caption{\textbf{Energy conversion.}
    Time evolution of kinetic energy $KE(t)$ and released potential
    energy $\delta PE(t)$ (Eq.~\ref{eq:deltaP}) for $\Wdns$ and the
    held-out DNS test case $\Sim_5$ in the downsampled $128^3$ setting.
    $\Wdns$ captures the coupled conversion of gravitational potential
    energy into kinetic energy over the rollout window.}
    \label{fig:energy_budget}
\end{figure}

A complementary view of fidelity comes from the global energy budget. We
evaluate the volume integrated kinetic energy,
\begin{equation}
KE(t) = \int_V \frac{1}{2}\,\rho\,|\mathbf{v}|^2\,dV,
\end{equation}
and the released potential energy,
\begin{equation}
\delta PE(t)
=
(\rho_{1}-\rho_{2})\,g
\int_V \big[c(x,y,z,0)-c(x,y,z,t)\big]\, z \, dV,
\label{eq:deltaP}
\end{equation}
which is equivalent under our conventions to $PE(0)-PE(t)$. Figure~\ref{fig:energy_budget} shows that $\Wdns$ closely tracks the
DNS evolution of both $KE(t)$ and $\delta PE(t)$ over the rollout
window. The agreement is coupled in the physically expected way: as
mixing develops and gravitational potential energy is released, kinetic
energy rises in tandem. In other words, the model captures not only
the magnitude of these global quantities, but the energy-conversion
pathway that underlies RTI evolution. The close agreement in both
$KE(t)$ and $\delta PE(t)$ over the full rollout therefore reinforces
the picture already established by morphology, spectra, and late-time
growth: $\Wdns$ captures the large-scale dynamics of RTI in a
physically faithful way.

\subsection{Sample efficiency}
\label{sec:sample_efficiency}
 
\begin{figure}[htb!]
    \centering
    \includegraphics[width=\linewidth]{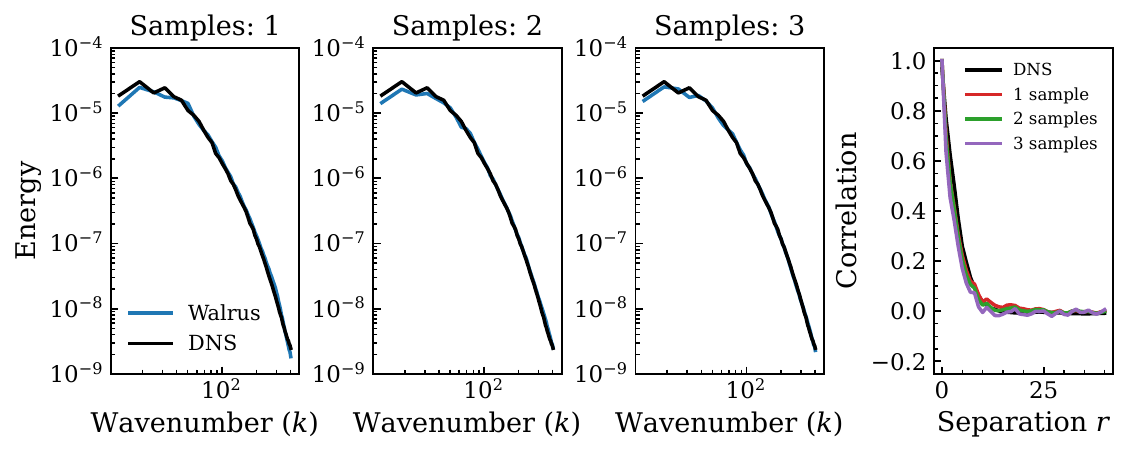}
    \caption{\textbf{Sample efficiency of 3D RTI finetuning.}
First three panels: $z$-averaged kinetic-energy spectra in the
$x$-$y$ planes of the held-out test case $\Sim_5$, computed at a
matched rollout time of $t=70$ in the self-similar regime for three independent
finetuning runs starting from the same pretrained checkpoint $\Wpre$
and using 1, 2, or 3 training realizations from
$\{\Sim_1,\Sim_2,\Sim_3\}$. In each case, the spectrum of the resulting
finetuned model is compared with the corresponding downsampled DNS
reference. Fourth panel: $z$-averaged radial two-point concentration
correlation function $C(r)$ on the same held-out test case,
where $r$ is the radial separation between point pairs in 
the $x$-$y$ plane. Even a single training realization recovers the spectral shape and
spatial correlation structure of the DNS reference. Additional training
realizations further improve spectral agreement, mainly for the highest wavenumbers.}
    \label{fig:sample_efficiency}
\end{figure}

We ask how much 3D RTI data are needed to specialize $\Wpre$. As before, the evaluation is performed on the fixed held-out test case, $\Sim_5$. We run three 
independent finetuning experiments, each starting from $\Wpre$ but 
using one, two, or all three of $\{\Sim_1,\Sim_2,\Sim_3\}$, and 
evaluate each on $\Sim_5$.

Figure~\ref{fig:sample_efficiency} addresses this question through 
two diagnostics: the kinetic-energy spectrum of $\Sim_5$ at $t=70$ 
in the self-similar regime, and the $z$-averaged radial two-point 
concentration correlation function. These probe complementary aspects 
of the held-out flow: the spectral distribution of kinetic energy and 
the spatial structure of the mixing field.

One training realization already captures the main large-scale physical character of the held-out flow. Finetuning on a single 3D RTI realization yields a model whose
predicted spectrum lies close to the DNS reference across the resolved
range, and whose concentration correlation function captures the
large-scale organization of the mixing layer. Adding further
realizations tightens the spectral agreement, with the clearest effect
appearing at the higher resolved wavenumbers, while the dominant
large-scale structure is already well captured from one sample. A
quantitative summary is given in
Appendix~\ref{app:sample_efficiency_metrics}.

The emulation results above show that a broad pretrained fluid model 
can be specialized into a physically faithful emulator of 
three-dimensional RTI turbulence with very little task-specific data. 
Luo et al.~\cite{luo2024fno} trained a task-specific Fourier neural 
operator on 300 simulated cases at $32\times32\times64$ resolution. 
Our study operates at $128^3$, more than thirty times larger in grid 
points, using one to three training realizations, and recovers the 
physical diagnostics by which RTI is judged over the full evolution 
of the instability.
\section{Emergent Behavior}
\label{sec:ood}

The emulation results establish that $\Wdns$ has learned 
physically faithful RTI dynamics from as few as one to three 
DNS realizations, recovering the canonical diagnostics of 
self-similar RTI growth across held-out simulations. We now 
ask whether the representations of RTI physics learned from 
finetuning $\Wpre$ on idealized DNS can transfer to settings 
the model was never trained on. The central question is 
sim-to-real transfer: can a model finetuned only on idealized 
DNS generalise to real laboratory conditions, without any 
experimental training? The successful emulation results raise 
a further question: if the model has learned the underlying 
physics of RTI, can it respond correctly in a physically 
distinct regime that demands a correct representation of 
buoyancy, given RTI is primarily a buoyancy-driven flow? We test this by initializing the finetuned model 
from stably stratified conditions, where a stable density 
gradient acts as a restoring force that suppresses and 
eventually arrests the mixing layer rather than driving it, 
a regime entirely absent from training.
 
\subsection*{Why physical RTI experiments resist simulation}
\label{subsec:why_resist_ic}

\begin{figure}[htb!]
    \centering
    \includegraphics[width=0.6\linewidth]{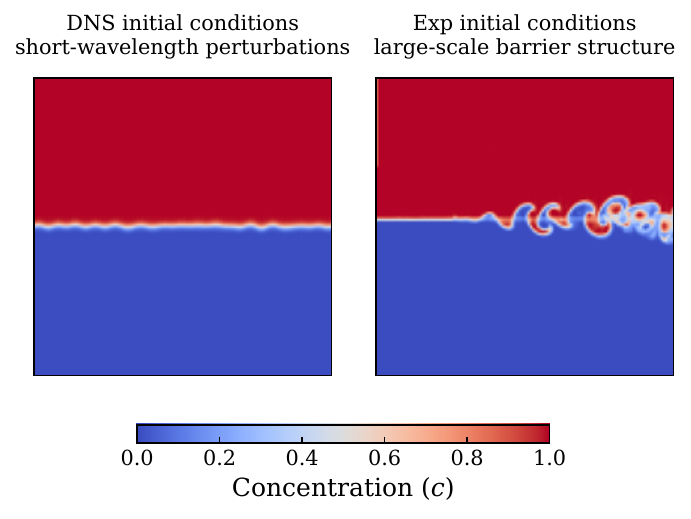}
    \caption{\textbf{Initial conditions: DNS vs. sliding-barrier 
    experiment.} Concentration fields for a representative DNS 
    realization (left) and a sliding-barrier experimental sample 
    (right). The DNS interface carries short-wavelength 
    perturbations characteristic of idealized numerical 
    initialization, whereas the experimental interface carries 
    large-scale structure set by barrier motion, structural 
    vibration, and molecular diffusion at release. This difference 
    in initial perturbation structure is a leading candidate 
    explanation for the sim-experiment discrepancy in $\alpha$ 
    discussed in Sec.~\ref{sec:intro}.}
    \label{fig:ic_comparison}
\end{figure}

Faithfully reproducing sliding-barrier RTI experiments in DNS 
remains an open 
problem~\cite{DALZIEL1993127,dalziel1999selfsimilar}. In a 
sliding-barrier apparatus, the initial perturbation spectrum is 
set by motions from the barrier 
removal, structural 
vibration, and molecular diffusion at the interface, none of 
which are precisely known or easily 
parametrized, and the large-scale 
velocity field at release carries measurement noise and systematic 
drift absent from idealized DNS (Appendix~\ref{app:experimental_methods}) ~\cite{DALZIEL1993127,dalziel1999selfsimilar}. The contrast in initial perturbation structure between 
DNS and experiment is illustrated in Fig.~\ref{fig:ic_comparison}. While both laboratory flows and 
idealized simulations operate at comparable Atwood numbers 
($A_t$) within the Boussinesq regime, they otherwise occupy 
vastly different parameter spaces: real flows typically operate 
at much higher Schmidt numbers ($Sc \sim O(10^3)$) than standard 
DNS ($Sc \sim O(1)$), and evolve from far more complex initial 
conditions. Initialising DNS with nominal experimental parameters 
yields improved agreement with the late-time $\alpha$ measured 
in the 
laboratory~\cite{ramaprabhu_dimonte_andrews2005_initialperturbations,
zhou2017a_physrep_rtrm_i,zhou2017b_physrep_rtrm_ii}, but an 
exact match remains elusive.

Despite these stark differences in initial conditions and 
diffusivity, classical theory dictates that both systems 
ultimately cascade into the same self-similar, fully developed 
turbulent mixing regime (Sec.~\ref{sec:background}), governed 
by the same underlying physics. The value of $\alpha$ within 
that regime, however, depends on the initial conditions that seeded the flow \cite{DALZIEL1993127,dalziel1999selfsimilar}. Because this late-stage macroscopic evolution 
relies on the fundamental physics of self-similarity, we 
hypothesize that a finetuned foundation model that has correctly learned 
this physics from DNS will possess the representation necessary 
to correctly propagate to the experimental distribution. 

This is a non-trivial hypothesis. The model must 
have learned a deep representation of self-similar RTI 
physics, not one that can only interpolate within the distribution of the training data. Moreover, it must have learned how initial 
condition structure connects to the late-time value of $\alpha$ 
robustly enough to transfer from the short-wavelength, idealized 
perturbations of DNS to the large-scale, complex, 
apparatus-driven initial conditions of real laboratory experiments, a setting qualitatively unlike anything in the training data. That a model trained 
on a handful of idealized simulations could achieve this 
would be a direct demonstration of one of the central 
promises of foundation models for scientific 
ML~\cite{Channing_2026}: that representations learned 
from simulation can carry over to the complexity of the 
real world.

The emulation results of Sec.~\ref{sec:results_emulation} are encouraging: $\Wdns$, 
finetuned on a handful of DNS realizations, recovers the 
canonical physics diagnostics of self-similar RTI growth on 
held-out simulations with initial conditions statistically 
independent of the training set, suggesting the model has 
learned the underlying physics rather than memorized specific 
realizations. The next natural question
is whether that learned structure is robust enough to transfer to the
far more complex initial conditions of real laboratory experiments. This sim-to-real test is especially sharp because, as described in
Sec.~\ref{sec:intro}, most laboratory experiments report
$\alpha \approx 0.07$~\cite{read1984,snider1994,ramaprabhu2004,2009PhFl...21c4103O,youngs1989_physicad,DimonteSchneider1996,dalziel1999selfsimilar,mueschke2009_pof,Glimm2013,AndrewsDalziel2010},
whereas idealized miscible simulations more typically lie near
$\alpha \approx 0.02$-$0.03$~\cite{youngs2017_ps,dimonte2004alphagroup,cabotcook2006reynolds,youngs1994_lpb,linden1994molecular,cook2004mixingtransition,burton2011,Statsenko2013,livescu2011dns_rti,Livescu2010,Youngs2003,Youngs2013}. We therefore focus
on late-time $\alpha$ as our primary experimental diagnostic, since
this is where the experiment--simulation gap is most clearly defined
and where the success of sim-to-real transfer can be assessed most
directly. Accordingly, all experimental transfer claims in this section
are made at the level of the late-time growth diagnostic, not at the
level of pointwise reconstruction of the full experimental flow.

\subsubsection*{Sim-to-Real: Setup}
\label{subsec:exp_setup}

The experimental corpus consists of six 2D planar slices 
derived from 3D laboratory RTI measurements in a sliding-barrier 
setup~\cite{dalziel1999selfsimilar,DALZIEL1993127,
nixon_thesis_placeholder}, an exceptionally sparse dataset. We 
denote these by 
$\Exp^{2D}=\{\Exp_1,\Exp_2,\Exp_3,\Exp_4,\Exp_5,\Exp_6\}$;
further details on the experimental methods are in 
Appendix~\ref{app:experimental_methods}. The experimental 
measurements are planar slices of the flow, so we finetune on 
2D planar slices extracted from one of our 3D DNS realizations, 
matching the dimensionality of the experimental data. We 
extract 10 such slices at different $y$ locations from $\Sim_1$ 
(see Sec.~\ref{sec:FT_walrus_on_RTI}). As before, we start with $\Wpre$ and finetune on the 2D DNS slices 
obtain $\WdnsTwoD$ (see Appendix~\ref{app:exp_finetuning} for 
details), specialising the model to RTI physics in the 2D 
setting without using any experimental data. The choice to use 
a single 3D realization for creating 2D slices is guided by the rapid saturation 
observed in Sec.~\ref{sec:sample_efficiency}. We use an input 
context of $L=2$ frames, capturing the onset of the 
barrier-removal phase without exposing the model to its full 
duration (${\sim}\,4\,\mathrm{s}$); sensitivity to this choice 
is examined in Appendix~\ref{app:context_ablation}.

Unlike the clean, statistically homogeneous DNS fields the 
model was trained on, the experimental slices carry measurement 
noise, unknown initial perturbation spectra, non-periodic 
boundary effects, and apparatus-specific artefacts. The 
setting, sparse lower-dimensional experimental measurements of 
an inherently three-dimensional flow, is not unique to RTI. 
Many laboratory diagnostics capture planar or point-wise 
snapshots while the true flow is three-dimensional, and 
experimental datasets are rarely large. The approach described 
here may generalise to such settings.

\subsection*{Zero-shot behavior on experimental data}
\label{subsec:sim-to-real}

\begin{figure}[htb!]
    \centering
    \includegraphics[width=\linewidth]{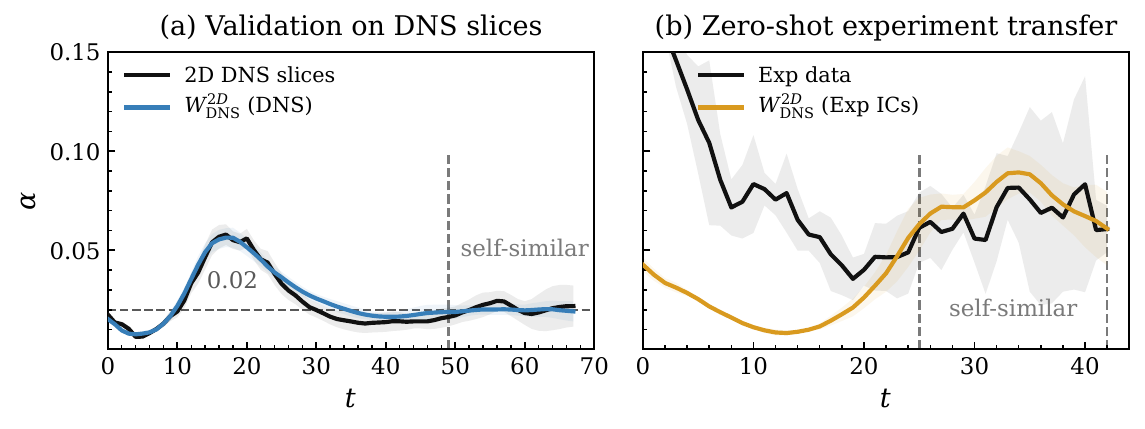}
\caption{\textbf{Zero-shot transfer from idealized DNS to experimental RTI data.}
\textbf{(a)} Validation on held-out 2D DNS slices. The black curve shows the mean growth coefficient $\alpha(t)$ computed over the held-out 2D DNS slices, with the shaded band denoting the spread across slices. The blue curve, labeled $\WdnsTwoD(\mathrm{DNS})$, shows the corresponding rollout of the DNS-specialized model $\WdnsTwoD$ initialized from DNS input frames. The horizontal dashed line marks $\alpha=0.02$, and the vertical dashed marker indicates the approximate onset of the late-time DNS self-similar regime. In the self-similar regime ($t \gtrsim 50$), black (DNS reference sim) and blue (Walrus rollout) both converge to the idealized DNS $\alpha$ value $\sim$ 0.02.
\textbf{(b)} Zero-shot transfer to held-out experimental RTI data. The black curve shows the experimental reference samples (labeled Exp data) with shaded bands denoting one standard deviation, while the gold curve shows $\WdnsTwoD$ initialized from experimental frames, labeled $\WdnsTwoD(\mathrm{Exp~ICs})$, without any experimental finetuning. The dashed vertical markers denote the late-time experimental comparison window, beginning at $t\sim25$, where the experimental flow is approximately self-similar. In this window, the zero-shot rollout rises into the same high-$\alpha$ growth band as the held-out experiments. Thus, the same DNS-specialized model, with the same neural network weights, reaches a different late-time growth regime depending only on the input: DNS frames lead to the low-$\alpha$ DNS regime, while experimental frames lead to the higher-$\alpha$ experimental regime. Early-time agreement in panel (b) is not expected because the sliding-barrier transient is absent from the DNS training data and only the first two experimental frames are supplied as input. The critical point is that the same model, $\WdnsTwoD$, produces qualitatively different late-time behavior at inference time depending
solely on whether the input frames come from simulation or experiment, demonstrating successful sim-to-real transfer.}
\label{fig:exp_zeroshot}
\end{figure}

$\WdnsTwoD$, despite \textit{never} having seen experimental 
data, enters the late-time growth band observed in the 
laboratory. In the approximately self-similar window beginning 
at $t \sim 25$ (Fig.~\ref{fig:exp_zeroshot}b), the model's 
rollout on experimental initial frames rises into the same 
$\alpha$ band as the held-out experimental samples. We confirmed 
this requires RTI finetuning: $\Wpre$, run on the same 
experimental frames without any RTI specialization, produces 
unstable rollouts (Appendix~\ref{app:trivial_alternative}). This 
rules out the possibility that any smooth propagator initialized 
from experimental frames would drift toward the experimental 
late-time regime; RTI specialization from DNS is essential.

At earlier times, the evolution is shaped by the sliding-barrier 
release, and the model does not track the experimental trajectory 
closely (Fig.~\ref{fig:exp_zeroshot}b). This is expected. $\WdnsTwoD$ sees only the first two 
experimental frames, which carry a limited imprint of the 
barrier-driven transient, and the DNS slices it was trained on 
contain no barrier release at all. The model has therefore not 
learned how the full transient unfolds. It is also worth noting 
that the experimental flow enters its approximately self-similar 
regime earlier than idealized DNS (Fig.~\ref{fig:exp_zeroshot}a vs. Fig.~\ref{fig:exp_zeroshot}b): the large-scale initial 
structure introduced by the barrier release drives rapid growth 
from the outset, shortening the transient compared to idealized 
DNS where large-scale structure must develop through nonlinear 
mode coupling and bubble 
merger~\cite{DALZIEL1993127,dalziel1999selfsimilar,
ramaprabhu_dimonte_andrews2005_initialperturbations}. Crucially, 
this early-time mismatch does not prevent correct late-time 
behavior. Classical theory holds that regardless of how the 
transient unfolds, both systems ultimately cascade into the same 
self-similar regime (Sec.~\ref{sec:background}). The model, 
initialized from experimental frames carrying large-scale 
initial structure, produces a late-time $\alpha$ consistent with 
the experimental regime, suggesting that this initial structure 
plays a role in setting the late-time growth rate.

The key result is most clearly seen by comparing two rollouts of the same model, shown in Fig.~\ref{fig:exp_zeroshot}. When $\WdnsTwoD$ is initialized from 2D DNS frames (blue), it settles from about $t \gtrsim 50$ into a plateau near $\alpha \approx 0.02$ (Fig. \ref{fig:exp_zeroshot}a), the expected late-time value for idealized DNS. When initialized from experimental frames instead (gold), the same model with the same neural network weights enters the higher-$\alpha$ regime (Fig. \ref{fig:exp_zeroshot}b), rising into the band occupied by the real experimental data (black) in the approximately self-similar window. The only difference between these two rollouts is the input (Fig.~\ref{fig:ic_comparison}): one starts from clean DNS frames, the other from real experimental frames carrying the large-scale initial structure set by the barrier release. This contrast reveals two things about what the model has learned. First, it suggests that the model has learned the physics of self-similarity: given DNS-like initial 
conditions, it produces DNS-like late-time growth (blue line). Second, that learned physics is not tied to the specific DNS training distribution. When given experimental initial frames carrying large-scale structure absent from all training data, the model extrapolates correctly to a different self-similar regime, producing a higher late-time $\alpha$ consistent with those conditions (gold line). The
model was trained only on DNS with short-wavelength initial
conditions and low $\alpha$, yet it responds correctly to
qualitatively different input by producing qualitatively
different output. This suggests that the model has learned more than the DNS-specific
realization of self-similar RTI growth. It has learned a dependence of
late-time growth on the structure of the initial condition, general
enough to move from the DNS-like $\alpha$ regime to the experimental
one when initialized from laboratory frames. The same story 
is visible in the averaged concentration field: Appendix~\ref{app:cbr_zero_shot} shows space-time maps of the 
horizontally averaged concentration profile $\bar{c}(t,z)$
for both rollouts, where the zero-shot experimental rollout 
develops a broader mean profile at late times, consistent
with a larger mixed region and the upward shift in $\alpha$.

To confirm this result is not tied to the particular choice of $L=2$, we trained three separate DNS-specialized 2D models starting from $\Wpre$
using context lengths $L=1$, $L=2$, and $L=3$, and evaluated 
each zero-shot on the same held-out experimental samples 
(Appendix~\ref{app:context_ablation}). All three show the same 
upward shift in late-time $\alpha$. The $L=1$ case is especially 
notable: the model sees only a single experimental frame, i.e. the initial condition, with 
no direct information about how the barrier release is 
unfolding, yet still predicts late-time growth well above the 
DNS-like $\alpha \approx 0.02$ and toward the experimental 
self-similar band. The shift toward the experimental regime is 
therefore a robust feature of the learned representation, not an artifact of how many input frames are provided.

This result bears on the decades-long debate in the RTI literature 
over the origin of the experiment-simulation discrepancy in 
$\alpha$, as described in Sec.~\ref{sec:intro}. The fact that 
experimental early-time input alone is sufficient to move 
$\WdnsTwoD$ into the experimentally observed late-time growth 
regime supports the view that this discrepancy is driven, to a 
substantial degree, by the conditions established during the initial 
experimental 
release~\cite{2005PhPl...12e6301D,
ramaprabhu_dimonte_andrews2005_initialperturbations,
youngs2017_ps,zhou2017a_physrep_rtrm_i}. Crucially, this 
evidence is independent of and complementary to purely numerical 
approaches: rather than modifying the simulation to better 
reproduce the experiment, we ask what a data-driven model learns 
when exposed to real experimental conditions for the first time. 
That it crosses into the experimental regime purely from seeing real initial frames provides evidence that initial conditions 
play a substantial role in driving the discrepancy. 

Foundation model transfer thus offers a new and independent perspective on a problem that has resisted
resolution through simulation alone. More broadly, these
results suggest that physics foundation models can
transfer the learned physics from idealized simulation to
real experimental conditions, even when the two occupy
different regimes. For physical systems where laboratory
conditions resist faithful numerical capture, this is an
encouraging result for the broader programme of sim-to-real transfer in scientific machine learning.

\subsection*{Finetuning on real-world experimental data}
\label{subsec:ft_real_data}

\begin{figure}[htb!]
    \centering
    \includegraphics[width=\linewidth]{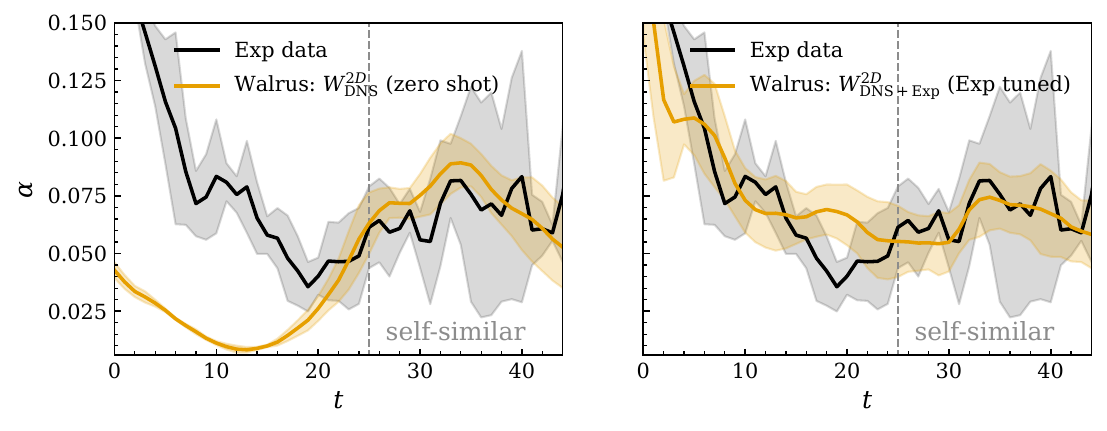}
    \caption{\textbf{$\alpha(t)$ for held-out test set experimental 
    data: zero-shot transfer and experimentally adapted predictions.}
    Left: rollout of $\WdnsTwoD$ on the three held-out experimental 
    samples, without any experimental finetuning. Right: rollout of 
    $\Wexp$, obtained after a second finetuning stage on the two 
    remaining experimental samples and evaluated on the same three 
    held-out cases. In both panels, black denotes the experimental 
    data and gold the Walrus prediction. The vertical dashed line 
    marks the onset of the late-time comparison window. Experimental 
    finetuning mainly improves agreement through the transient and 
    intermediate stages of the evolution, while preserving the larger 
    late-time $\alpha$ values already reached by $\WdnsTwoD$ at 
    zero-shot. Solid lines show the mean across rollouts; shaded 
    regions indicate $\pm 1$ s.d.}
    \label{fig:exp_tuned}
\end{figure}

The zero-shot result establishes that $\WdnsTwoD$ can enter 
the late-time experimental growth regime without any 
experimental training. A natural follow-up question is how 
much finetuning on a small amount of experimental data can further improve 
the model. This tests a key promise of the 
pretraining-finetuning paradigm: that a model with a strong 
simulation-trained prior requires very little experimental 
data to adapt toward real laboratory conditions.

Specialising 
to RTI using simulation alone first, then making a minimal 
adjustment with only two experimental samples, lets us 
separately assess what the model learns from simulation and 
what further finetuning on experimental data adds. We 
therefore split $\Exp^{2D}$ into two samples for this 
finetuning stage, one for validation, and three for testing, 
the same three held-out cases used to evaluate $\WdnsTwoD$ 
at zero-shot. Finetuning $\WdnsTwoD$ on the two experimental 
training samples yields $\Wexp$, evaluated on the same three 
held-out cases at input context $L=2$.

Experimental finetuning improves agreement through the earlier 
and intermediate stages of the experimental evolution 
(Fig.~\ref{fig:exp_tuned}), while the late-time growth regime, 
already reached at zero-shot, is largely preserved. The main 
improvement is in the transient: in zero-shot, the model sees 
only the first two experimental frames and does not learn how 
the full barrier-driven transient unfolds; experimental 
finetuning exposes it to that structure, substantially reducing 
the early-time mismatch. The concentration field evolution in 
Appendix~\ref{app:exp_ic} tells the same story 
(Fig.~\ref{fig:exp_ic_rollout}): $\Wexp$ qualitatively captures the 
barrier-release-driven early-time structure, which is notoriously hard to do with DNS.

This result has a broader scientific implication. As discussed 
in Sec.~\ref{subsec:why_resist_ic}, the initial conditions of sliding-barrier 
experiments have proven to be persistently difficult to be captured with DNS: barrier motion, structural vibration, and 
molecular diffusion all contribute to the perturbation spectrum 
at release, and none are precisely 
known~\cite{DALZIEL1993127,dalziel1999selfsimilar,
ramaprabhu_dimonte_andrews2005_initialperturbations}. Prior 
numerical approaches addressed this by building experimentally 
measured initial conditions directly into the 
simulation~\cite{dalziel1999selfsimilar,
ramaprabhu_dimonte_andrews2005_initialperturbations,
mueschke2009_pof,youngs2017_ps}, requiring bespoke numerical 
codes tailored to each experimental setup. Here, two experimental 
samples suffice for the model to qualitatively learn this complex IC structure 
in a purely data-driven way. Analysis of the dominant coherent structures in the early-time 
vorticity field using dynamic mode decomposition (DMD) \cite{schmid2010dmd} shows 
that $\Wexp$ shifts toward the broad anisotropic large-scale 
structure of the experimental release, away from the regular 
columnar structure of the idealized DNS 
(Fig.~\ref{fig:early_time_dmd}, Appendix~\ref{app:exp_ic}). The model does not need to know why the initial conditions are complex; finetuning on two samples are enough to capture the dominant structure 
of the experimental transient in a purely data-driven way.

The second finetuning stage uses a learning rate of 
$2\times10^{-6}$, 50 times smaller than the $10^{-4}$ used to 
obtain $\WdnsTwoD$ (see Appendix~\ref{app:exp_finetuning}). 
Generating rollout predictions takes about 20-30 seconds after 
finetuning on a single H100 GPU. RTI laboratory experiments take 
days to weeks of human effort to 
generate~\cite{DALZIEL1993127,dalziel1999selfsimilar,nixon_thesis_placeholder}, and faithful DNS 
reproductions of experimental conditions remain 
difficult~\cite{dalziel1999selfsimilar,
ramaprabhu_dimonte_andrews2005_initialperturbations}. Two 
experimental samples and a lightweight finetuning stage suffice 
to bring a simulation-trained model into quantitative agreement 
with real laboratory samples. For physical systems where both 
experimental data and faithful DNS are hard to come by, this 
suggests that lightly adapting a simulation-trained foundation 
model offers a promising, purely data-driven route to capturing 
the complex conditions of real laboratory experiments. 

Together with the zero-shot result, this demonstrates that the 
pretraining-finetuning paradigm can bridge the gap between 
idealized simulation and real laboratory conditions with 
remarkably little data, suggesting it as a promising approach 
for physics problems where both experimental data and faithful 
simulation are scarce.

\subsection*{Zero-shot transfer to stably stratified RTI}
\label{subsec:zero-shot-stably-stratified}

Beyond the sim-to-real transfer, the emulation results 
of Sec.~\ref{sec:results_emulation} raise a further 
question regarding zero-shot transfer: if the model has learned the underlying physics 
of RTI, can it respond correctly in a physically distinct 
regime that demands a correct representation of buoyancy, 
given that RTI is fundamentally a buoyancy-driven flow? To probe that, We return 
here to $\Wdns$, the model finetuned on 3D DNS in 
Sec.~\ref{sec:results_emulation}, and initialize it from 
conditions with stable background stratification at zero-shot, where a 
stable density gradient acts as a restoring force that 
suppresses and eventually arrests the mixing layer rather than driving it, a
regime entirely absent from
all finetuning data, and evaluate its zero-shot behavior.

\begin{figure}[htb!]
    \centering
    \includegraphics[width=\linewidth]{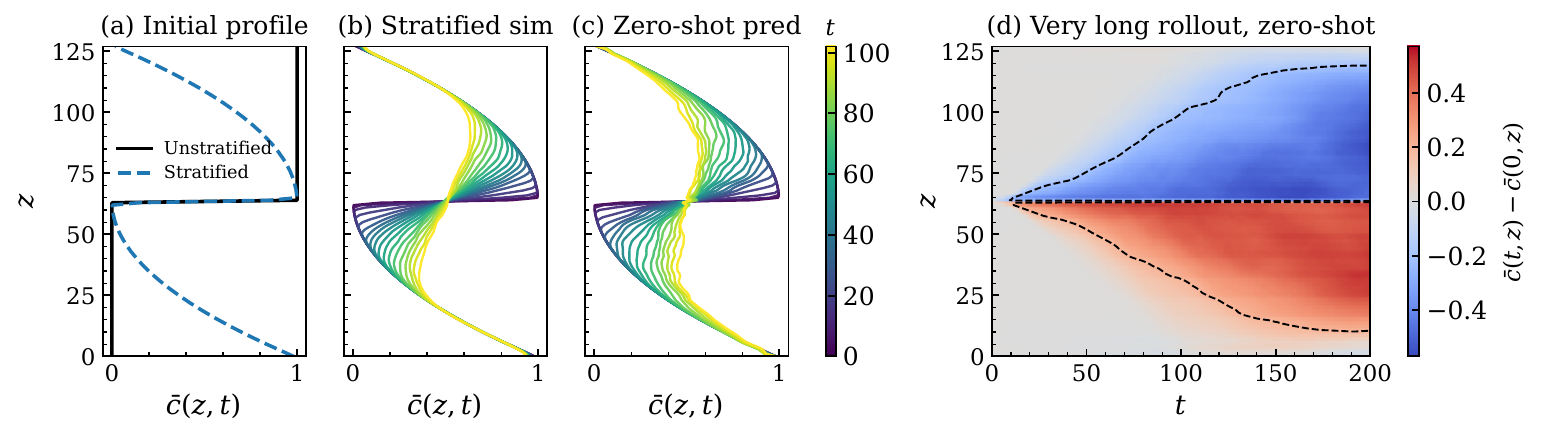}
    \caption{\textbf{Zero-shot transfer of Walrus finetuned on 
    unstratified DNS to stratified Rayleigh-Taylor flow.} 
    \textbf{(a)}~Initial horizontally averaged concentration 
    profiles for the unstratified and stratified cases. 
    \textbf{(b)}~Evolution of the mean concentration 
    profile in the stratified reference DNS. 
    \textbf{(c)}~Corresponding zero-shot rollout from $\Wdns$, 
    finetuned only on unstratified RTI DNS. Stable stratification 
    confines the partially mixed region near the midplane rather 
    than allowing the broader spreading of the unstratified case. 
    $\Wdns$ captures this qualitative shift without ever seeing 
    stratified flows during training, though it predicts slightly 
    weaker confinement at late times. \textbf{(d)}~Very long 
    zero-shot rollout of $\Wdns$ out to 200 timesteps, showing 
    continued deceleration of the mixing layer beyond the 
    training horizon.}
    \label{fig:stratified_vs_non_stratified}
\end{figure}

The effect of stratification in the reference simulation is unmistakable (panel b of Fig. \ref{fig:stratified_vs_non_stratified}). Mixing remains confined around the midplane,
and the mean concentration profiles do not relax into the broader spreading characteristic of the unstratified case, in
line with prior studies showing that stable background stratification suppresses vertical spreading and confines RTI-
driven mixing \cite{Lawrie_Dalziel_2011,davieswykesdalziel2014_jfm}. Walrus reproduces this qualitative response in zero shot (panel c of Fig. \ref{fig:stratified_vs_non_stratified}). Although it was specialized only on
unstratified RTI, its predicted profiles remain similarly confined and do not revert to the unstratified regime. 
Fig.~\ref{fig:stratified_vs_non_stratified} (d) extends the 
zero-shot rollout to 200 timesteps. The mixing layer growth continues 
to decelerate throughout, consistent with physical 
expectation~\cite{davieswykesdalziel2014_jfm}, even though the 
training trajectories ended at 100 timesteps.

The discrepancy is one of degree rather than kind: $\Wdns$ 
predicts somewhat weaker confinement at late times than the 
reference simulation, but the qualitative response to 
stratification is correct, suggesting the model has encoded a physical understanding of buoyancy-driven mixing in its learned representation which it can extend to unseen physical regimes.

Together with the sim-to-real results, these findings 
suggest a broader conclusion : the physics encoded during 
finetuning on idealized simulation can generalize well beyond the specific training distribution. It can carry over to the 
noise and complexity of real laboratory conditions, and 
extend to physical regimes the model has never seen. For 
physical systems where faithful simulation is difficult 
and experimental data is scarce, this is an encouraging 
sign that foundation models offer a promising paradigm for applications in the physical sciences.
 
\section{Conclusion}
\label{sec:conclusion}

This paper asks whether physics foundation models finetuned on 
idealized simulation can transfer to real laboratory conditions 
and to qualitatively new physical regimes. For RTI, the answer 
is yes. Finetuned on a small number of DNS realizations and never 
shown a single experimental frame, Walrus \cite{mccabe2025walrus} enters the late-time 
growth regime observed in real RTI laboratory experiments. A model 
finetuned on the same DNS but never exposed to stratified flows 
transfers zero-shot to stably stratified RTI, a physically 
different regime absent from all training data. In both cases, 
the learned representation carries physical structure well outside 
the simulation manifold used for finetuning. The sim-to-real transfer therefore provides new, independent, and
purely data-driven evidence that initial conditions play a critical
role in the longstanding sim-experiment discrepancy in
\(\alpha\)~\cite{2005PhPl...12e6301D,
ramaprabhu_dimonte_andrews2005_initialperturbations,
youngs2017_ps,zhou2017a_physrep_rtrm_i}.

RTI was chosen as a compelling system precisely because it 
combines physical significance, resistance to standard ML methods, 
and a well-quantified simulation-experiment discrepancy. The 
results suggest that the pretraining-finetuning paradigm offers 
a promising route to sim-to-real transfer in physical systems 
where laboratory data is sparse, expensive to obtain, and 
difficult to capture in idealized simulation, reducing the 
dependence on large experimental datasets and bespoke numerical 
reproductions of laboratory conditions.

Several limitations bound these conclusions. The DNS used here are $256^3$ in resolution, below the resolution of the highest-fidelity RTI 
benchmarks~\cite{livescu2011dns_rti}. The experimental evaluation 
is limited to two-dimensional planar fields; whether the agreement 
in late-time $\alpha$ extends to fully three-dimensional 
experimental RTI remains to be tested. Recent work pushes ML 
surrogates to $512^3$ for 3D homogeneous isotropic 
turbulence~\cite{holzschuh2026pd}, suggesting resolution 
constraints will ease as the computational landscape matures.

Until we can determine \textit{how} Walrus represents the flow 
well enough to transfer across the sim-to-real divide, its 
inferences remain a fiction that emulates Rayleigh-Taylor 
instability without traditional fluid dynamical modelling. Yet 
Walrus captures the essence of DNS and experiments well beyond 
its training. Traditional simulations and the Navier-Stokes 
equations are themselves fictions, approximations of true physics, 
just as experimental measurements, though arising from true 
physics, are only approximate representations. The question is 
whether foundation models offer new, useful fictions despite 
relying on imperfect training data. Our results suggest they do.

\section{Acknowledgments}
We would like to acknowledge the support of Schmidt Sciences and the Simons Foundation. This work was supported in part by the AI2050 program at Schmidt Sciences (Grant G-25-70028). Dr. Mukhopadhyay thanks the Infosys-Cambridge AI centre for support. Additionally, computations were run at facilities supported by the Scientific Computing Core at the Flatiron
Institute. The Flatiron Institute is a division of the Simons Foundation. The authors thank Lucy Reading-Ikkanda for assistance with figures. Miles Cranmer is grateful for support from the Schmidt Sciences AI2050 Early Career
Fellowship and the Isaac Newton Trust. S.S. Nixon and R. Watteaux thank the CEA's Centre de Calcul Recherche et Technologie for facilitating DNS computations. S. S. Nixon gratefully acknowledges CEA for funding his PhD research and thanks the technical staff at the G. K. Batchelor Laboratory (DAMTP, University of Cambridge) for their invaluable assistance in completing the experiments.

\section*{Author Contributions}
Payel Mukhopadhyay, Stefan S. Nixon, Stuart B. Dalziel and Miles Cranmer conceived the study. Payel Mukhopadhyay designed, led, 
and carried out the machine learning development, 
finetuning experiments, and scientific analysis underlying the results presented in the paper. Payel Mukhopadhyay wrote the manuscript, with Stefan S. Nixon contributing to writing and revision. Michael McCabe and Payel Mukhopadhyay co-led the development of the 
pretraining model which was used as a starting point of the finetuning 
analyses of the paper. Stefan S. Nixon and Stuart B. 
Dalziel designed and conducted the laboratory RTI 
experiments and collected the experimental data. 
Romain Watteaux and Stefan S. Nixon provided the direct numerical 
simulations. Payel Mukhopadhyay, Stefan S. Nixon, 
Miles Cranmer, Stuart B. Dalziel, and Romain Watteaux 
contributed to the core scientific discussion and refinement 
of the manuscript. Contributions of the remaining authors range from broader feedback on the paper to compute assistance for the project.

\clearpage
\bibliographystyle{unsrt}
\bibliography{refs}
\clearpage
\onecolumn
\section{Appendix}

\subsection{Pretraining of the checkpoint $\Wpre$}
\label{app:pretraining_wpre}

The pretrained checkpoint $\Wpre$ used throughout this work follows the
Walrus pretraining recipe of~\cite{mccabe2025walrus} closely. Rather than repeating the full model-development paper here, we
summarize only the aspects most relevant to the present application and
refer the reader to~\cite{mccabe2025walrus} and its appendix for the
full architecture, distributed-training strategy, and complete
pretraining corpus. That corpus spans a broad mixture of 2D and 3D
continuum-dynamics datasets, including examples such as shear flows and
magnetohydrodynamic (MHD) turbulence from The Well dataset collection \cite{ohana2024the}. The checkpoint used here, however, was constructed with RTI excluded from pretraining. The list of datasets used in the pretraining is shown in Table \ref{tab:wpre_datasets}.

As described in Sec.~\ref{sec:FT_walrus_on_RTI}, Walrus is trained as a
next-step delta-prediction model. In pretraining, the model uses a
temporal context length of 6 for 2D datasets and 3 for 3D datasets, reflecting
the memory budget of the joint 2D/3D training recipe in
\cite{mccabe2025walrus}. The pretraining objective is the mean
absolute error (MAE) loss, matching the loss used in the Walrus paper
and in our downstream finetuning.

The checkpoint $\Wpre$ was trained on a broad mixture of two- and
three-dimensional continuum-dynamics datasets through the Walrus / The
Well data pipeline. It uses the same 1.3B-parameter Walrus architecture
family as in~\cite{mccabe2025walrus}: an isotropic space-time
transformer with adaptive encoder/decoder tokenization, causal temporal processing, and joint support for both 2D and 3D data. The model is trained with a unified field index map spanning 67 physical state variables. Full architectural details are given in the Walrus paper~\cite{mccabe2025walrus}; the dataset list relevant to the present checkpoint is summarized in Table~\ref{tab:wpre_datasets}.

Optimization follows the Walrus pretraining recipe of
~\cite{mccabe2025walrus} closely and the reader is encouraged to read that paper for more details. We use AdamW with learning rate
$3.3\times10^{-4}$, together with the same learning-rate schedule,
mixed-data loading strategy, and distributed-training setup as in the
Walrus paper. The context length is 6 for 2D datasets and 3 for 3D
datasets, again matching the original pretraining recipe. Note that these pretraining context lengths should be distinguished from the
downstream RTI finetuning choices used in this paper: the 3D emulation
results use \(L=3\), whereas the main 2D experimental-transfer results
use \(L=2\), with \(L=1\) and \(L=3\) examined in the additional ablation studies of
Appendix~\ref{app:context_ablation}.

The most important point for the present paper is conceptual rather than
architectural: $\Wpre$ is a broad fluid checkpoint that has learned from
diverse continuum-dynamics data, but it was trained with RTI excluded
from pretraining. The downstream results in this work therefore probe
whether that broad prior can be specialized efficiently to RTI and still
retain physically meaningful transfer beyond the finetuning set.

\begin{table*}[htb!]
\centering
\caption{\textbf{Datasets used in pretraining of $\Wpre$.}
The pretraining corpus follows the Walrus recipe~\cite{mccabe2025walrus}
closely, but with Rayleigh-Taylor instability excluded. The table
therefore mirrors the broad 2D/3D continuum-dynamics mixture used in
Walrus, while omitting the RTI dataset.}
\label{tab:wpre_datasets}
\small
\begin{tabular}{llllccc}
\hline
Dataset & Short name & Source & Coordinates & Resolution & $n_{\mathrm{steps}}$ & $n_{\mathrm{traj}}$ \\
\hline
acoustic\_scattering\_discontinuous      & Discontinuous    & The Well & $(x,y)$       & $256\times256$       & 102  & 2000 \\
acoustic\_scattering\_inclusions         & Inclusions       & The Well & $(x,y)$       & $256\times256$       & 102  & 4000 \\
acoustic\_scattering\_maze               & Maze             & The Well & $(x,y)$       & $256\times256$       & 202  & 2000 \\
active\_matter                           & Active Matter    & The Well & $(x,y)$       & $256\times256$       & 81   & 360 \\
euler\_multiquadrants\_periodicBC        & MultiQuadrantsP  & The Well & $(x,y)$       & $512\times512$       & 101  & 5000 \\
euler\_multiquadrants\_openBC            & MultiQuadrantsO  & The Well & $(x,y)$       & $512\times512$       & 101  & 5000 \\
gray\_scott\_reaction\_diffusion         & Gray-Scott      & The Well & $(x,y)$       & $128\times128$       & 1001 & 1200 \\
helmholtz\_staircase                     & Staircase        & The Well & $(x,y)$       & $1024\times256$      & 50   & 512 \\
MHD                                      & MHD (3D)         & The Well & $(x,y,z)$     & $64\times64\times64$               & 100  & 100 \\
planetswe                                & PlanetSWE        & The Well & $(\theta,\phi)$ & $256\times512$     & 1008 & 120 \\
rayleigh\_benard                         & Rayleigh-Benard & The Well & $(x,y)$       & $512\times128$       & 200  & 1750 \\
shear\_flow                              & Shear Flow       & The Well & $(x,y)$       & $256\times512$       & 200  & 1120 \\
supernova\_explosion                     & Supernova (3D)   & The Well & $(x,y,z)$     & $128\times128\times128$              & 59   & 1000 \\
turbulence\_gravity\_cooling             & TGC (3D)         & The Well & $(x,y,z)$     & $64\times64\times64$ & 50   & 2700 \\
turbulent\_radiative\_layer\_2D          & TRL (2D)         & The Well & $(x,y)$       & $128\times384$       & 101  & 90 \\
turbulent\_radiative\_layer\_3D          & TRL (3D)         & The Well & $(x,y,z)$     & $128\times128\times256$ & 101 & 90 \\
viscoelastic\_instability                & Viscoelastics    & The Well & $(x,y)$       & $512\times512$       & variable & 260 \\
FPOHarmonics                             & FBHarmonics      & FlowBench & $(x,y)$       & $512\times128$       & 242  & 400 \\
\hline
\end{tabular}
\end{table*}


\subsection{Initial Perturbations for DNS}\label{app:Initial_Conditons}

The initial interface displacement, $\eta(\mathbf{x})$, is constructed as a random-phase, narrow-band perturbation to ensure the Rayleigh-Taylor instability begins deeply within the linear regime. The geometric and dynamic characteristics of the perturbation field are governed by three dimensionless parameters:

\begin{itemize}
    \item \textbf{Perturbation Reynolds Number ($Re$):} Represents the ratio of inertial to viscous forces at the initially excited length scale. Physically, a high $Re$ drives the system into a highly non-linear, chaotic regime at late times, which induces small temporal decorrelation timescales, a critical consideration for the predictive horizon of sequence-based ML models. In practice, this is defined using the Atwood number ($A_t$), gravitational acceleration ($g$), and the kinematic viscosity ($\nu$) as:
    \begin{equation}
        Re = \frac{\sqrt{A_t g k_0}}{\nu k_0^2}
    \end{equation}
    
    \item \textbf{Spectral Bandwidth ($B$):} Defines the width of the excited spectrum ($\Delta k$) relative to the mean wavenumber, given by $B = \Delta k / k_0$. Physically, $B$ determines the multiplicity of distinct length scales present in the initial perturbation. This directly influences the duration and dynamics of the transient phase before the system transitions into a fully developed, Kolmogorov-like turbulent inertial cascade.
    
    \item \textbf{Initial Steepness ($S$):} Dictates the physical root-mean-square amplitude ($\eta_0$) relative to the mean wavenumber ($k_0$), defined as $S = k_0 \eta_0$. A value of $S \ll 1$ ensures the perturbation does not trigger premature non-linear mode coupling.
\end{itemize}
The physical displacement field is generated via the frequency domain. A two-dimensional wavenumber spectrum, $\hat{\eta}(\mathbf{k})$ with magnitude $K = |\mathbf{k}|$, is populated exclusively within the spectral annulus bounded by the mean wavenumber and the bandwidth
\begin{equation}
    k_0 - \frac{\Delta k}{2} \le K \le k_0 + \frac{\Delta k}{2}.
\end{equation}

Modes within this annulus are assigned a uniform amplitude and a random phase $\phi(\mathbf{k}) \in [0, 2\pi)$ to yield unbiased, isotropic noise. Conjugate symmetry, $\hat{\eta}(\mathbf{k}) = \hat{\eta}^*(-\mathbf{k})$, is strictly enforced to guarantee a purely real physical domain. The field is brought into physical space via an inverse Fourier transform
\begin{equation}
    \eta(\mathbf{x}) = \sum_{\mathbf{k}} \hat{\eta}(\mathbf{k}) e^{i \mathbf{k} \cdot \mathbf{x}}.
\end{equation}

To investigate the instability across different spatial resolutions, and specifically to ensure the resulting datasets could be rigorously compared over similar temporal rollout lengths for machine learning prediction, the initial perturbation parameters were tailored so both configurations transition into a fully developed, self-similar turbulent regime over a similar physical time-frame. For the $256^3$ grid resolution, the initial field was parameterized with $\{Re, B, S\} = \{7, 0.3, 0.1\}$, corresponding to a mean spectral wavenumber of $k_0 = 32$ and a root-mean-square (RMS) amplitude of $\eta_0 = 5 \times 10^{-4}$. Conversely, for the $128^3$ simulations, the parameters were adjusted to $\{Re, B, S\} = \{5, 0.7, 0.12\}$, which centers the spectrum at $k_0 = 20$ and sets the RMS amplitude to $\eta_0 = 1 \times 10^{-3}$. While the general methodology and perturbation magnitudes remain comparable to the foundational $\alpha$-group study by Dimonte \textit{et al.}~\cite{dimonte2004alphagroup}, these targeted adjustments to the mean wavenumber, bandwidth, and steepness were necessary to compensate for resolution-dependent effects. Ultimately, this allowed for a temporally consistent evolution into late-time mixing beneficial for training and evaluating the foundation model's efficacy.


\subsection{3D RTI finetuning used to obtain $\Wdns$}
\label{app:finetuning_3d}

The 3D RTI model $\Wdns$ is obtained by finetuning the pretrained
checkpoint $\Wpre$ on the downsampled RTI dataset described in
Sec.~\ref{sec:FT_walrus_on_RTI}. The five DNS realizations
$\Sim^{3D}=\{\Sim_1,\Sim_2,\Sim_3,\Sim_4,\Sim_5\}$ are first
block-averaged from native $256^3$ resolution to an effective $128^3$
representation. As discussed in Sec.~\ref{sec:results:morphology}, this
choice is deliberate: the aim is to train the model on RTI dynamics
represented at $128^3$ as sampled from a higher-fidelity flow, rather
than on the detailed local structure of a native low-resolution
realization.

For the main 3D emulation results, we use $\Sim_1,\Sim_2,\Sim_3$ for
training, $\Sim_4$ for validation, and $\Sim_5$ as the held-out test
case. Each realization uses a different random initialization of the
perturbation spectrum, so the validation and test cases are
statistically independent realizations of the same physical regime. All
sample-efficiency experiments reported in the main text follow the same
logic: they are separate finetuning runs that start again from the same
pretrained checkpoint $\Wpre$, but use only one, two, or three training
realizations, while keeping the same held-out test case $\Sim_5$.

The finetuning objective is the same next-step delta-prediction task
introduced in Sec.~\ref{sec:FT_walrus_on_RTI}. The model receives
\(L=3\) consecutive input states and predicts the next-step increment.
Training uses a fixed temporal stride of 1 simulation step. The loss is
again MAE, as in pretraining. Input states consist of concentration and
nondimensionalized velocity fields, with concentration defined from
density as in Eq.~(6) and velocity rescaled by the free-fall velocity
scale \(\sqrt{A_t g H}\). 

Optimization follows the Walrus finetuning recipe
of~\cite{mccabe2025walrus} closely and the reader is encouraged to read that paper for more details. Starting from $\Wpre$, all 3D
finetuning runs use AdamW with learning rate \(3\times10^{-4}\) and
weight decay \(10^{-4}\), together with the same inverse-square-root schedule as ~\cite{mccabe2025walrus}. The total finetuning budget is 400K samples, compared with 500K samples
in~\cite{mccabe2025walrus}. This keeps task-specific adaptation below the original Walrus setting rather than increasing the budget for this task. This fine-tuning is performed on four H100 GPUs.

Checkpoint selection for $\Wdns$ is based on performance on the validation realization $\Sim_4$. In particular, among saved checkpoints we retain the one with the lowest validation error in the global
kinetic and potential energy evolution relative to the ground truth, rather than selecting purely by pointwise loss. This choice aligns model selection with the physically meaningful diagnostics emphasized in the main text, especially the energetic quantities discussed in Sec.~\ref{sec:results:energy}, including the kinetic energy \(KE(t)\) and released potential energy \(\delta PE(t)\) from Eq.~\ref{eq:deltaP}. The resulting checkpoint defines $\Wdns$ and is
used for all 3D emulation results reported in Sec.~\ref{sec:results_emulation}. The DNS realization reported in the main text correspond to the test DNS $\Sim_5$ which was not used during the optimization or validation process.

\subsection{Native $128^3$ RTI finetuning}
\label{app:128x3_native}

We also consider a separate 3D finetuning setting in
which the pretrained checkpoint $\Wpre$ is finetuned directly on native
$128^3$ RTI simulations. We denote the resulting model by
$\WdnsNative$. These native-$128^3$ simulations follow the same
physical setup as the downsampled case used in the main text: they
consist of five statistically independent RTI realizations with
randomly initialized perturbation spectra, of which three are used for
training, one for validation, and one as the held-out test case. This
native-$128^3$ model is therefore distinct from $\Wdns$, which in the
main text is trained and evaluated in the downsampled
$256^3 \rightarrow 128^3$ setting. We include the native-$128^3$
results here to show that direct finetuning at native resolution can
yield strong pointwise and local reconstruction. That, however,
is not the primary setting of the study. The main text instead
emphasizes the downsampled $256^3 \rightarrow 128^3$ construction
because it more cleanly probes whether the model learns RTI dynamics at
$128^3$ as sampled from a higher-fidelity flow, rather than the
detailed local structure of any particular realization.

\begin{figure}[htb!]
    \centering
    \includegraphics[width=0.8\linewidth]{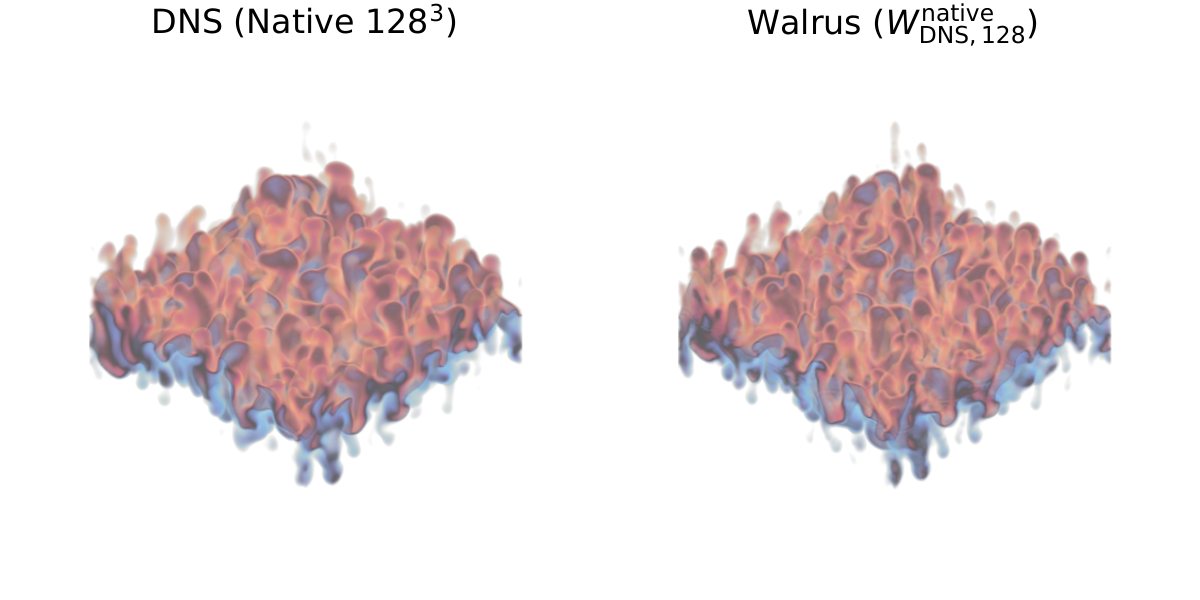}
    \caption{\textbf{Qualitative agreement for $\WdnsNative$ at native $128^3$ resolution and $t=60$.}
Comparison between the held-out native $128^3$ DNS and the
corresponding prediction of $\WdnsNative$, obtained by finetuning
$\Wpre$ directly on native $128^3$ RTI data. This setting is separate
from the downsampled $256^3 \rightarrow 128^3$ setup used in the main
text. Direct native-$128^3$ finetuning yields good local
reconstruction while again preserving the large-scale bubble morphology
and mixed-layer structure of the held-out case.}
    \label{fig:native_RTI}
\end{figure}

\begin{figure}[htb!]
    \centering
    \includegraphics[width=0.8\linewidth]{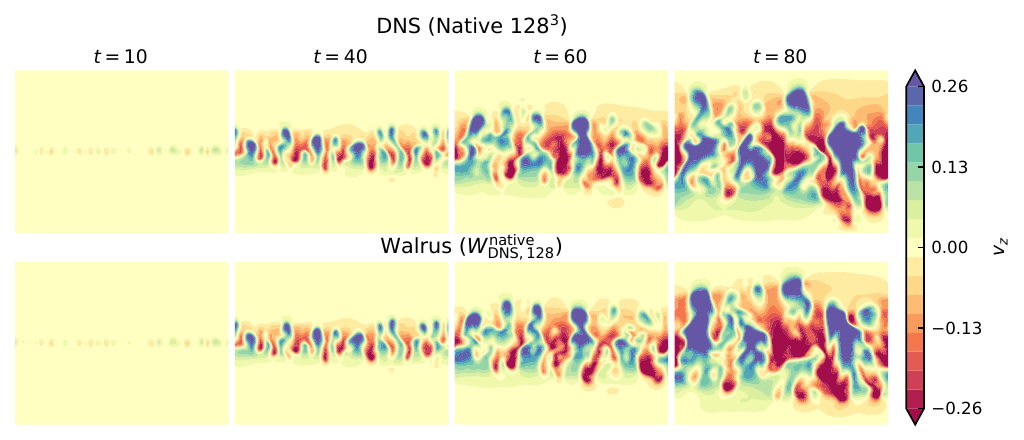}
    \caption{\textbf{Rayleigh-Taylor flow structure for $\WdnsNative$ at native $128^3$ resolution.}
Vertical cross-sections of the vertical velocity component $v_z$ at
multiple rollout times for the held-out native $128^3$ direct numerical
simulation (DNS, top row) and the corresponding prediction of
$\WdnsNative$ (bottom row). The slices correspond to an $x$-$z$ plane
taken at the midplane of the domain in the $y$ direction. In this native-$128^3$ setting, $\WdnsNative$
captures local vertical-velocity structures of the mixing layer with high qualitative fidelity.}
\label{fig:native_RTI_contours}
\end{figure}

Figure~\ref{fig:native_RTI} shows a representative comparison at
$t=60$. Also in this native-$128^3$ setting, $\WdnsNative$ captures
the large-scale bubble morphology and the overall mixed-layer structure
of the held-out simulation with high qualitative fidelity.
Figure~\ref{fig:native_RTI_contours} shows the corresponding evolution
of vertical velocity structure across time. Here the agreement is also good at the local level: the model captures the dominant
sign-structured \(v_z\) features that organize the plume dynamics. In
that sense, native-$128^3$ finetuning provides a useful complementary
view of model behavior. It shows that when training and evaluation are
performed directly at the same native resolution, Walrus can achieve strong pointwise agreement deep into the rollout,
even though the main paper deliberately focuses on the more physically
controlled downsampled setting.

\clearpage

\subsection{Breakdown of standard ML baselines on the emulation task of 3D RTI dynamics}
\label{app:baseline_breakdown}

We show representative autoregressive rollouts from
three standard baseline architectures used in The Well
benchmark~\cite{ohana2024the}: the Fourier Neural Operator (FNO) \cite{li2021fourier}, the
Tensorized Fourier Neural Operator (TFNO) \cite{kossaifi2024multigrid}, and ConvNeXt-UNet \cite{liu2022convnet}. These baselines are present as a part of the Well codebase and we use the model configurations present in the Well. Additionally, we tune the learning rate between $10^{-3}, 10^{-4}, 5\times 10^{-3}$ to find the optimal learning rate for each model and show the best results possible. We also make sure to give the same compute budget to these baselines models as given to $\Wdns$, i.e. these models see the same number of training samples (400K) as used for $\Wdns$. These models are strong and widely used baselines in scientific machine learning, and they perform well on many PDE surrogate tasks \cite{ohana2024the}. For the three-dimensional RTI setting studied here, however, all three break down qualitatively under rollout.

Rather than sustaining the evolution of a growing mixed layer with coherent
RTI bubble structure, the rollouts develop grossly unphysical
artifacts: spurious large-scale structures that overwhelm the true instability
dynamics. The problem is therefore structural. The models do not
preserve the physically meaningful three-dimensional morphology of the
flow over time.

This qualitative breakdown is consistent with the broader argument of the study. RTI is difficult because it is a chaotic prediction
problem and success must be judged by whether a model
preserves the coupled interface geometry, spectral cascade, and global evolution of the mixing layer over long autoregressive rollouts. The figures below show that standard task-specific baselines do not do so
reliably for the RTI emulation problem studied here.

\begin{figure}[htb!]
    \centering
    \includegraphics[width=\linewidth]{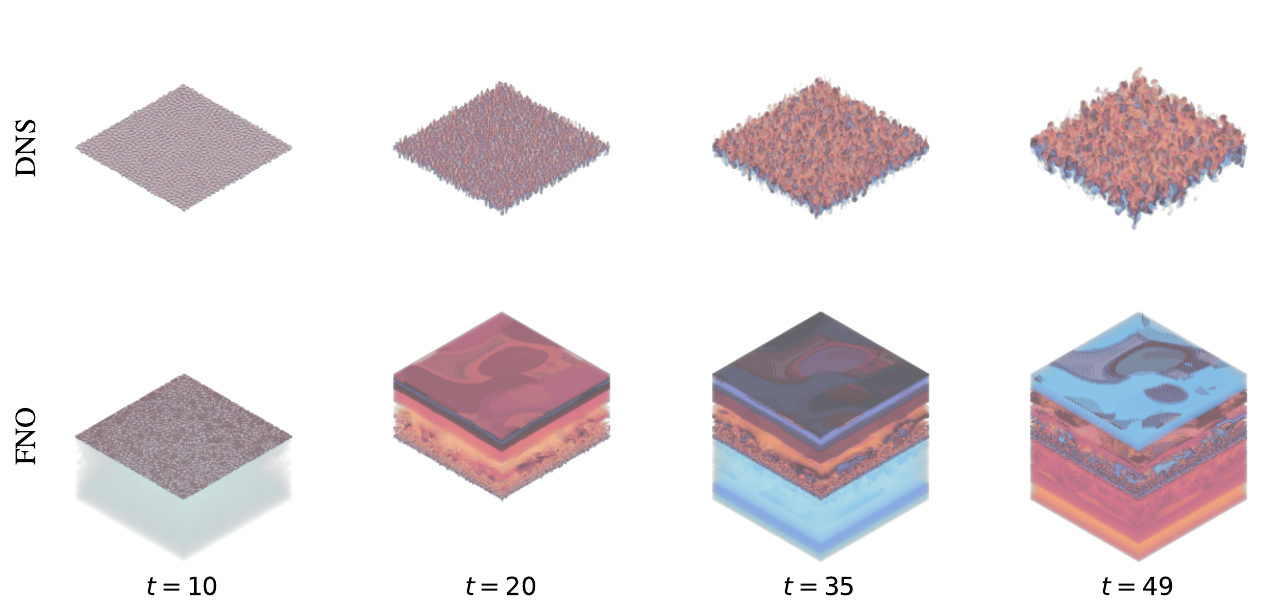}
    \caption{\textbf{Breakdown of the Fourier Neural Operator (FNO) on 3D RTI rollout.}
    Comparison between the held-out DNS trajectory (top row) and the
    corresponding autoregressive FNO rollout (bottom row) at four
    representative times. While the DNS develops the expected RTI mixed
    layer and bubble morphology, the FNO prediction quickly departs from
    physically plausible evolution and forms large slab-like structures
    that do not resemble RTI dynamics. The failure is therefore not
    just quantitative, but qualitative: the model does not preserve the
    three-dimensional flow structure of the instability over time.}
    \label{fig:baseline_fno_breakdown}
\end{figure}

\begin{figure}[htb!]
    \centering
    \includegraphics[width=\linewidth]{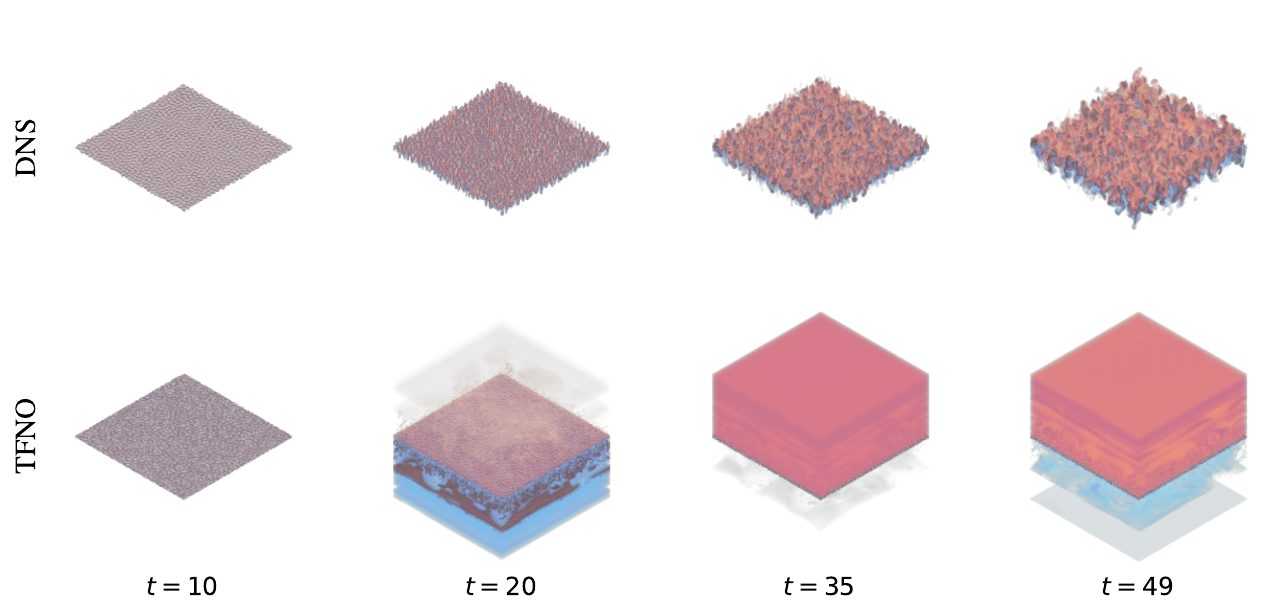}
    \caption{\textbf{Breakdown of the Tensorized Fourier Neural Operator (TFNO) on 3D RTI rollout.}
    Comparison between the held-out DNS trajectory (top row) and the
    corresponding autoregressive TFNO rollout (bottom row) at four
    representative times. After an initially plausible state, the TFNO
    rollout collapses into highly unphysical artifacts and fails to sustain the development of a realistic RTI mixed layer.
    This illustrates that standard operator-learning baselines can
    struggle severely with long-rollout emulation of 3D RTI, even at a qualitative level.}
    \label{fig:baseline_tfno_breakdown}
\end{figure}

\begin{figure}[htb!]
    \centering
    \includegraphics[width=\linewidth]{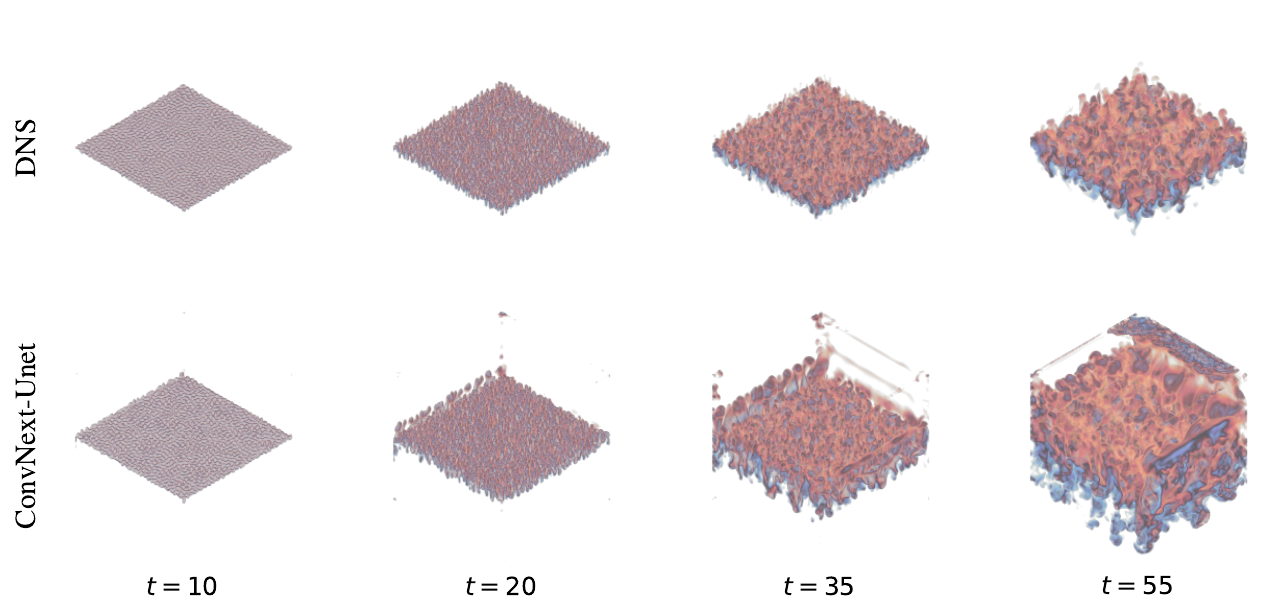}
    \caption{\textbf{Breakdown of ConvNeXt-UNet on 3D RTI rollout.}
    Comparison between the held-out DNS trajectory (top row) and the
    corresponding autoregressive ConvNeXt-UNet rollout (bottom row) at
    four representative times. Although the early state remains roughly
    plausible, the rollout soon develops spurious artifacts that are incompatible with the expected RTI
    evolution.}
    \label{fig:baseline_convnext_breakdown}
\end{figure}

\clearpage
\subsection{Robustness to patch-jittering}
\label{app:robust_stochastic}

\begin{figure}[htb!]
    \centering
    \includegraphics[width=0.8\linewidth]{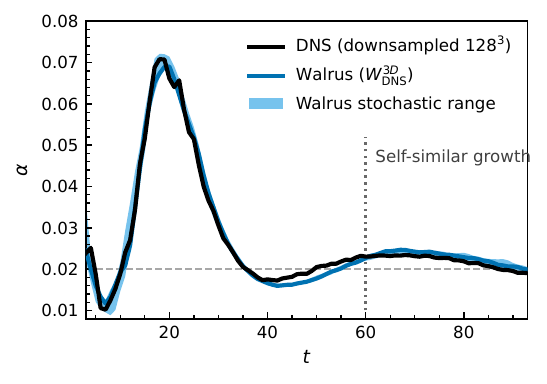}
    \caption{Evolution of the $\alpha$ value for DNS and $\Wdns$, shown along with a shade of blue band which shows the randomness associated to patch jittering. The blue band is created by taking 10 predicted outputs of $W_{\mathrm{DNS}}$ on the same initial conditions of the test set. }
    \label{fig:robust_jitter}
\end{figure}

As noted in the main text, Walrus architecture has some inherent randomness in its output predictions, which is a consequence of the patch jittering technique \cite{mccabe2025walrus}. Walrus is based on a vision transformer \cite{dosovitskiy2021an} based backbone, wherein the input frames are first converted into so-called `tokens" also known as ``patches" wherein images are first divided into non-overlapping patches of a fixed patch size, which is then embedded into a vector before being processed by the transformer. This patchification process is necessary to keep the number of tokens processed by the transformer under control, since compute scales quadratically with the token count. However, having a fixed patch size during autoregressive rollouts is a well known failure mode of vision transformers and is known to degrade rollout quality of transformers based Physics emulators \cite{mukhopadhyay2026overtone}. The patch-jittering technique, introduced in the Walrus paper solves this problem by randomly translating the patches during the training process thereby stabilizing long rollouts. A by-product is the addition of some run-by-run variation arising out of this. As shown in Fig. \ref{fig:robust_jitter}, the $\alpha$ prediction is robust to the randomness caused by the jitter (shown as a blue band). Additionally, we have checked that all of our emulation and zero-shot results shown in the paper are robust to this randomness.

\subsection{Band-averaged spectral error for the sample-efficiency study}
\label{app:sample_efficiency_metrics}

To quantify the spectral overlap in
Fig.~\ref{fig:sample_efficiency}, we compute a band-averaged error
between the Walrus prediction and the DNS reference for the held-out
test case \(\Sim_5\). Let \(E_{\mathrm{Walrus}}(k)\) denote the
\(z\)-averaged kinetic-energy spectrum of a Walrus rollout and let
\(E_{\mathrm{DNS}}(k)\) denote the corresponding DNS spectrum, both
evaluated at the same rollout time. For a prescribed band \(B\), we
define the mean absolute relative spectral error as
\begin{equation}
\varepsilon_B
=
\frac{1}{|B|}
\sum_{k \in B}
\left|
\frac{E_{\mathrm{Walrus}}(k)-E_{\mathrm{DNS}}(k)}
     {E_{\mathrm{DNS}}(k)}
\right|,
\label{eq:band_spectral_error}
\end{equation}
where \(|B|\) is the number of discrete shell-averaged spectral
wavenumbers in the band.

To reduce sensitivity to any single rollout frame, we average
\(\varepsilon_B\) over the late-time window \(t=70,\dots,80\), which
lies within the self-similar regime. For the quantitative summary, we
focus on the two informative ranges
\(k \in [16\pi,64\pi)\) and \(k \in [64\pi,128\pi]\), which correspond
to intermediate and smallest resolved scales in the shell-averaged
spectrum. 

The resulting values support the qualitative picture in the main text.
A single training realization already captures the dominant large-scale
structure of the held-out flow well (as shown in Fig. \ref{fig:sample_efficiency}), while additional training
realizations tighten agreement over the smallest resolved scales (highest wavenumbers). This is also shown in Table \ref{tab:sample_efficiency}.

\begin{table}[htb!]
\centering
\caption{\textbf{Band-averaged spectral error for the sample-efficiency study.}
Mean absolute relative error \(\varepsilon_B\) from
Eq.~\ref{eq:band_spectral_error} between the Walrus and DNS
kinetic-energy spectra of the held-out test case \(\Sim_5\), averaged
over the late-time window \(t=70,\dots,80\) in the self-similar
regime.}
\label{tab:sample_efficiency_errors}
\small
\begin{tabular}{lcc}
\hline
Training samples & \(k \in [16\pi,64\pi)\) & \(k \in [64\pi,128\pi]\) \\
\hline
1 & 0.1069 & 0.2926 \\
2 & 0.0566 & 0.2034 \\
3 & 0.0422 & 0.0791 \\
\hline
\end{tabular}
\label{tab:sample_efficiency}
\end{table}


\subsection{Experimental Methods}\label{app:experimental_methods}

\begin{figure}[h]
    \centering
    \includegraphics[width=0.42\textwidth]{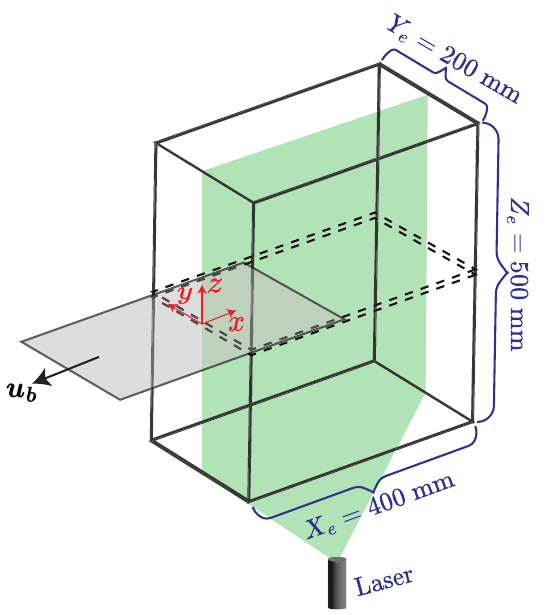}
    \caption{Experimental setup. The polycarbonate barrier (light grey) is removed at velocity $\vec{u}_b$.}
    \label{fig:tank}
\end{figure}

\paragraph{Apparatus:}
A sliding barrier apparatus initiates RTI in a stationary frame. A rectangular acrylic tank ($0.5\times0.4\times0.2~\si{\metre}$, coordinates $(X_e,Y_e,Z_e)$, $Z_e$ vertical) holds two fluid layers of differing density separated by a transparent polycarbonate barrier at $z=0$ (figure~\ref{fig:tank}). The polycarbonate allows full-domain imaging from the instant of removal, unlike earlier stainless-steel or composite designs ~\cite{DALZIEL1993127,dalziel1999selfsimilar}. Interior surfaces are matt-black coated to suppress stray reflections.

A dual-cavity Litron Nano~L~100 Nd:YAG laser ($\SI{50}{\milli\joule}$, $\SI{532}{\nano\metre}$, $\SI{10}{\nano\second}$ pulses, $\SI{100}{\hertz}$ per cavity) is shaped into a sheet $1$-$\SI{3}{\milli\metre}$ wide ~\cite{brown2018exploring}; a $\SI{30}{\second}$ warm-up stabilises pulse-to-pulse intensity ~\cite{grayson2018impact}. Two Allied Vision Bonito CMC-4000 cameras (4~MPixel, $100~\si{\hertz}$, synchronised via DigiFlow) record density (Camera~A: $f/1.2$, long-pass Raman filter) and velocity (Camera~B: $f/1.4$, $\SI{532}{\nano\metre}$ bandpass) simultaneously. The lower layer is brine (NaCl, $\mathit{Sc}\approx700$); the upper is brine-alcohol with refractive index matched to the lower ~\cite{DALZIEL1993127}.

\paragraph{Density and PLIF:}\label{sec:app_plif}
Rhodamine~6G (peak absorption ${\approx}\SI{527}{\nano\metre}$, added to the lower fluid) fluoresces proportionally to local concentration ~\cite{crimaldi2008planar}. Frames are processed by background subtraction, ray-parallel coordinate mapping (calibrated from a grid at the tank base, correcting for sheet divergence and refraction), wavelet-Fourier streak removal ~\cite{munch2009stripe}, and Beer-Lambert attenuation correction using uniform-dye calibration frames, then mapped back to lab coordinates. Validation against a motorised conductivity probe gives agreement within $0.8\%$ of the density range and ${\approx}50\%$ noise reduction over standard methods; the minimum sheet width (${\approx}\SI{1}{\milli\metre}$) is comparable to the Kolmogorov scale ~\cite{linden1994molecular}, confirming dissipation-range resolution.

\paragraph{Velocity and PIV:}\label{sec:app_piv}
Both layers are seeded with $\SI{50}{\micro\metre}$ polyamide particles ($\mathit{St}\ll1$). DigiFlow ~\cite{sveen2005dynamic,partridge2019versatile} processes image pairs using $6\times6$~px interrogation windows (${\approx}1.6\times1.6~\si{\milli\metre}^2$), sub-pixel cross-correlation, and polynomial coordinate calibration; the $1$-$\SI{3}{\milli\metre}$ sheet thickness retains most particles between frames.

\paragraph{Initial Conditions:}\label{sec:app_initial_conditions}
PIV from the instant of barrier removal reveals three perturbation features: an overturning circulation on the withdrawal side (driven by asymmetric void-filling, consistent with a vortex sheet shed from the trailing edge ~\cite{dalziel1999selfsimilar}); shear layers about $z=0$ that roll up into wall vortices; and a wake instability behind the retreating barrier. Weak convective cells driven by surface cooling are observed but secondary. The perturbation amplitude has little effect on the ultimate mixing extent ~\cite{dalziel1999selfsimilar}, though its spectral content influences early-time mode growth.

\paragraph{Error Sources:}\label{sec:app_error_sources}
Several error sources affect measurement fidelity. \emph{Sensor and illumination noise} — CCD read noise and shot-to-shot laser variability — introduce uncorrelated and multiplicative intensity errors in PLIF; these are mitigated by the warm-up period, high pulse energy, and low dye concentrations. \emph{Optical artefacts} from tank-base scratches produce vertical streaks removed by wavelet-Fourier filtering; the Raman edge filter's angle-dependent cut-on is non-material in the central analysis region, confirmed using a bandpass filter. \emph{Residual refractive-index mismatch} near the interface introduces minor errors in the PLIF mapping and PIV displacements, largest when the interface is sharp. \emph{Mechanical vibrations} from the laboratory floor produce spurious long-wavelength fluid motions, while \emph{surface cooling} drives weak convective cells in the upper layer; both are secondary at the Atwood numbers studied. \emph{Photobleaching and photoquenching} — irreversible dye degradation under repeated laser exposure and fluorescence suppression by Cl$^-$ collisional quenching, respectively — are managed by low dye concentrations, short run durations, and routine calibration. Finally, \emph{barrier-removal asymmetry} (variation in removal speed and seal condition) is the primary source of run-to-run variability in early-time RTI growth statistics.


\subsection{2D RTI finetuning used to obtain $\WdnsTwoD$ and $\Wexp$}
\label{app:exp_finetuning}

The 2D models used in the experimental-transfer part of the study are obtained in two stages. Starting from the pretrained checkpoint $\Wpre$, we first finetune on a simulated 2D RTI dataset \(\mathcal{S}^{2D}\), constructed from the single 3D DNS realization \(\Sim_1\), to obtain $\WdnsTwoD$. We then use $\WdnsTwoD$ as the initialization for a second finetuning stage on the experimental dataset \(\Exp^{2D}=\{\Exp_1,\Exp_2,\Exp_3,\Exp_4,\Exp_5,\Exp_6\}\), yielding $\Wexp$. The experimental dataset consists of 2D crops of the concentration and velocity fields extracted from an underlying 3D laboratory RTI flow.

\paragraph*{Construction of the simulated 2D dataset \(\mathcal{S}^{2D}\).}
The dataset \(\mathcal{S}^{2D}\) is constructed from \(x\)–\(z\) planar
slices of the native \(256^3\) RTI DNS described in
Sec.~\ref{sec:FT_walrus_on_RTI}. From each selected 3D realization, we
extract 10 such slices at different \(y\) locations, so that each 2D sample corresponds to an \(x\)–\(z\) plane through the flow at a
distinct transverse position. For each slice, we retain the
concentration field together with the two velocity components defined on
that plane, \(v_x\) and \(v_z\), and discard the out-of-plane component
\(v_y\). The native \(256\times256\) \(x\)–\(z\) slices are then
converted to \(128\times256\) by cropping one spatial direction, so
that the simulated 2D inputs have the same overall spatial resolution
as the experimental measurements. We adopt this construction so that
the DNS-to-experiment transfer problem is posed on the same image size
and with the same field content available in the laboratory data.

For the first 2D finetuning stage, the simulated 2D slices inherit the same train/validation/test logic as the 3D setup: the 10 slices extracted from a single DNS realization, \(\Sim_1\) to be used for training, the 10 slices extracted from \(\Sim_4\) are used for validation, and the 10 slices extracted from \(\Sim_5\) define the held-out simulated 2D test set. The choice to use slices from a single 3D training realization for the 2D training stage is guided by the rapid saturation observed in the 3D sample-efficiency results of Sec.~\ref{sec:sample_efficiency}.

\paragraph*{First-stage finetuning on simulated 2D RTI DNS slices.}
To obtain $\WdnsTwoD$, we finetune $\Wpre$ on the training subset of
\(\mathcal{S}^{2D}\) using the same next-step delta-prediction
formulation introduced in Sec.~\ref{sec:FT_walrus_on_RTI}, but with
context length \(L=2\). Thus the model receives two consecutive input
frames and predicts the increment to the next one. Optimization uses
AdamW with learning rate \(10^{-4}\), weight decay \(10^{-4}\), the
same inverse-square-root schedule used in Walrus~\cite{mccabe2025walrus},
and the same MAE loss used in pretraining and 3D finetuning. Training
is performed with batch size 1 for 20,000 optimizer steps (takes $\sim$ 10 hrs to train) on 4 H100 GPUs.

Checkpoint selection for $\WdnsTwoD$ is based on the late-time growth
behavior of the validation rollouts rather than on pointwise loss
alone. In particular, among saved checkpoints we retain the one whose
validation predictions on the slices from \(\Sim_4\) best match the
self-similar-stage \(\alpha\) behavior of the corresponding simulated
2D reference slices. This choice reflects the role of the 2D model in
the paper: its primary purpose is not generic short-horizon
reconstruction, but transfer of the growth-rate diagnostic into the
experimental regime.

\paragraph*{Second-stage finetuning on experimental data.}
The experimentally adapted model $\Wexp$ is obtained by starting from
$\WdnsTwoD$ and performing a second finetuning stage on the laboratory
dataset \(\Exp^{2D}\). As described in the main text, the six
experimental samples are split into two for training, one for
validation, and three for testing. The same next-step delta-prediction
task is used, again with context length \(L=2\) in the main results.
This second stage is intentionally lightweight: it uses a learning rate
of \(2\times10^{-6}\), 50X smaller than the
\(10^{-4}\) used in the first finetuning stage on simulated 2D slices, and is run for 5000 optimizer steps (takes $\sim$ 2.5 hrs to train) with batch size 1 on 4 H100 GPUs.
This choice is deliberate. The goal is to adapt the DNS-specialized
model to the experimental domain while preserving the RTI
representation already learned from simulation. The second stage
therefore serves as a light experimental adjustment of a
simulation-trained prior, rather than a wholesale retraining of the
model on a very small and noisy dataset. Unless otherwise noted, the
remaining optimization setup follows the same 2D finetuning recipe as
above.

The important point is conceptual. $\WdnsTwoD$ is first specialized to
RTI using only simulated 2D slices. $\Wexp$ then adjusts that
simulation-specialized prior using only two experimental samples. This
two-stage design is what allows the main text to separate direct
transfer from DNS to experiment from subsequent adaptation to the
experimental regime.
\clearpage

\subsection{$\WdnsTwoD$ predictions evaluated zero-shot 
on experimental initial conditions}
\label{app:cbr_zero_shot}
\begin{figure}[htb!]
    \centering
\includegraphics[width=\linewidth]{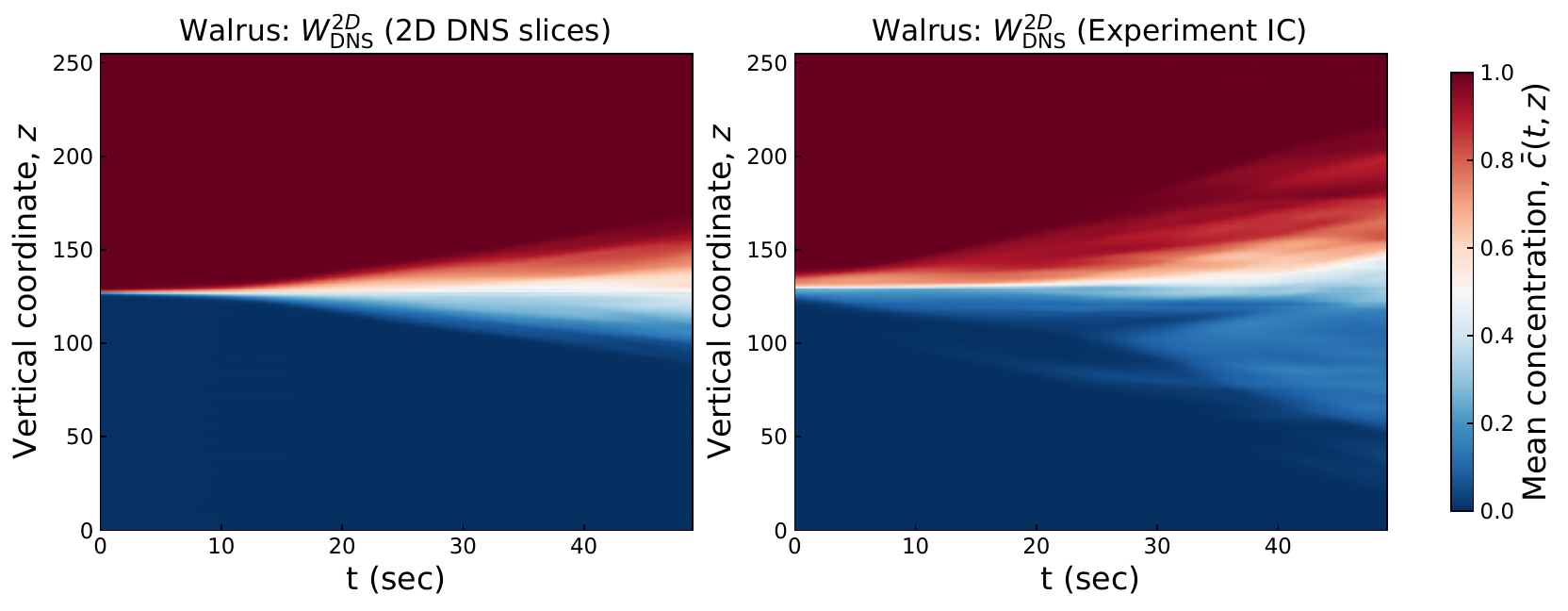}
    \caption{\textbf{Mean concentration evolution for 
    $\WdnsTwoD$ initialized from DNS and experimental 
    frames.} Space-time maps of the horizontally averaged 
    concentration profile $\bar{c}(t,z)$ (defined in 
    Sec.~\ref{sec:background}) for $\WdnsTwoD$ initialized 
    from 2D DNS frames (left) and the same model evaluated 
    zero-shot on experimental initial conditions (right). 
    Supplying experimental frames shifts the rollout away 
    from the DNS-like mixing regime toward broader late-time 
    mixing, consistent with the upward shift in $\alpha$ 
    shown in Fig.~\ref{fig:exp_zeroshot}.}
    \label{fig:walrus_cbar_zero_shot}
\end{figure}
To complement the late-time $\alpha(t)$ comparison in 
Sec.~\ref{subsec:sim-to-real}, we show the evolution of 
$\bar{c}(t,z)$ for $\WdnsTwoD$ initialized from 2D DNS 
frames and for the same model evaluated zero-shot on 
experimental initial conditions 
(Fig.~\ref{fig:walrus_cbar_zero_shot}). Over the plotted 
time window, the zero-shot experimental rollout develops 
a broader mean profile than the DNS-initialized rollout, 
consistent with a larger mixed region and the upward 
shift in late-time $\alpha$ discussed in 
Sec.~\ref{subsec:sim-to-real}.
\clearpage

\subsection{Robustness to variation of the number of input frames aka input temporal context}
\label{app:context_ablation}

\begin{figure*}[htb!]
\centering
\includegraphics[width=0.8\linewidth]{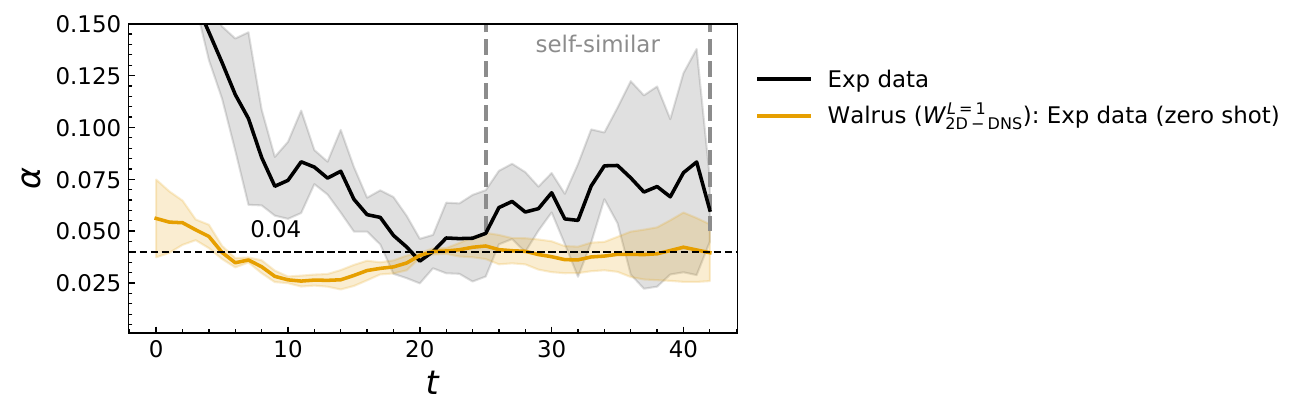}
\includegraphics[width=0.8\linewidth]{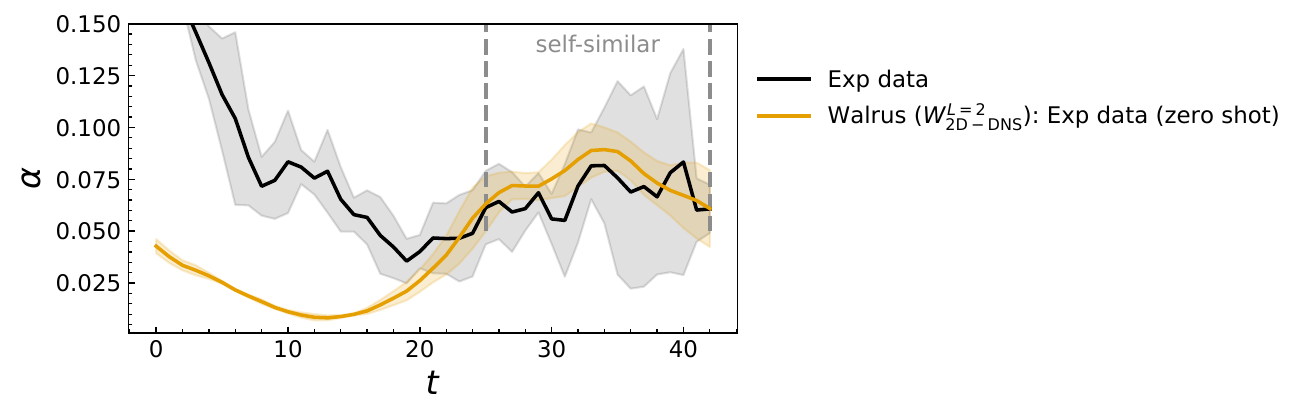}
\includegraphics[width=0.8\linewidth]{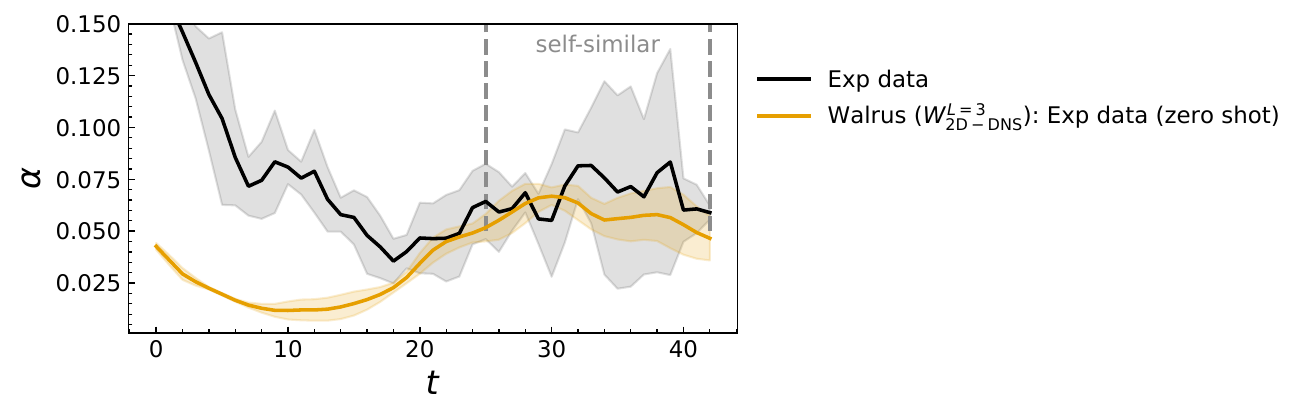}
\caption{\textbf{Robustness of zero-shot experimental transfer to the number of input frames.}
Zero-shot predictions on the held-out experimental samples for three
separate DNS-specialized 2D models, all initialized from $\Wpre$ and
then finetuned on simulated 2D RTI slices using context lengths
\(L=1\), \(L=2\), and \(L=3\), respectively. In each panel, black
denotes the experimental data and gold the corresponding Walrus
prediction on the experimental input without any experimental
finetuning. The dashed vertical lines mark the late-time window used
for comparison. The \(L=2\) setting is used in the main text. Even in
the particularly hard \(L=1\) setting, where the model is given only a single experimental frame as input, the predicted late-time growth moves well above the canonical DNS-like \(\alpha\approx0.02\) regime and toward
the experimental self-similar band.}
\label{fig:context_ablation}
\end{figure*}

In the main text, the zero-shot experimental-transfer results are shown
for context length \(L=2\). To test the robustness of that choice, we
train and evaluate three separate 2D DNS-specialized models. Starting
each time from the same pretrained checkpoint $\Wpre$, we finetune on
the simulated 2D RTI slice dataset using context lengths \(L=1\),
\(L=2\), and \(L=3\), yielding corresponding models
\(\WdnsTwoDL{1}\), \(\WdnsTwoDL{2}\), and
\(\WdnsTwoDL{3}\). Each model is then evaluated zero shot on the same
held-out experimental samples.

The \(L=1\) case is especially stringent. In that setting, the model is
given only a single experimental frame, so model only sees the very first initial experimental state alone, with no direct information about how the sliding-barrier release is unfolding. Notably, even under this harsh constraint, the late-time growth predicted by
\(\WdnsTwoDL{1}\), rises well above the canonical
DNS value \(\alpha\approx0.02\) and moves toward the experimentally observed self-similar regime. This is a compelling result showing that the increase in the late-time experimental $\alpha$ value is robust even under a constrained setting at zero-shot.

Increasing the context to \(L=2\) strengthens the result further. This
is the setting used in the main text, and it provides the clearest
agreement with the experimental late-time band while remaining a
demanding test: the model still sees only the very beginning of the
release phase (because of the two initial frames provided as input context), not its full evolution. The \(L=3\) case remains broadly
consistent with the same picture. Taken together, the three panels show
that the quantitative conclusion is stable across context lengths: once
$\WdnsTwoD$ is exposed to early experimental input, the resulting
zero-shot rollout leaves the low-\(\alpha\) DNS regime and moves toward the higher-\(\alpha\) growth behavior characteristic of the laboratory data.

This robustness matters for interpretation. It shows that the
zero-shot transfer reported in the main text is not tied to one
particular choice of the number of initial input frames shown to the model. Rather, the tendency of the
DNS-specialized model to respond to experimental input by entering a
higher late-time growth regime persists even when the amount of
input experimental information supplied at test time is varied substantially. 

\subsection{$\Wpre$ tested zero-shot on 2D DNS and experimental data, and corresponding rollouts of $\WdnsTwoD$}
\label{app:trivial_alternative}

The zero-shot experimental result in the main text would be much less
meaningful if it could be reproduced by any smooth autoregressive
propagator initialized from the experimental input. This appendix rules
out that trivial alternative. We compare the pretrained checkpoint
$\Wpre$, which has never been specialized to RTI, with the 2D
DNS-specialized model $\WdnsTwoD$.

\begin{figure}[htb!]
    \centering
    \includegraphics[width=\linewidth]{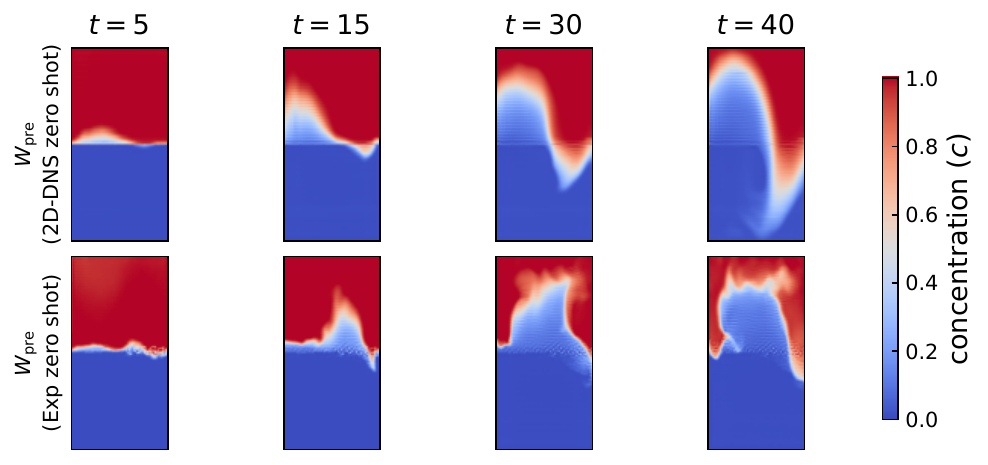}
    \caption{\textbf{Zero-shot rollouts of the pretrained model $\Wpre$ on 2D DNS and experimental input.}
    Top row: zero-shot rollout of $\Wpre$ initialized from 2D DNS
    input. Bottom row: zero-shot rollout of $\Wpre$ initialized from
    experimental input. Columns show representative times. In both
    cases, the pretrained checkpoint fails to produce a physically
    meaningful RTI evolution, showing that RTI behavior is not already
    present in $\Wpre$ and must be acquired through finetuning.}
    \label{fig:wpre_zero_shot_dns2d_exp}
\end{figure}

Figure~\ref{fig:wpre_zero_shot_dns2d_exp} shows zero-shot rollouts of
$\Wpre$ when initialized from 2D DNS input (top row) and from
experimental input (bottom row). In neither case does the pretrained
model produce a physically meaningful RTI evolution. On 2D DNS input,
it does not recover the expected RTI morphology. On experimental input,
it likewise fails to generate a coherent late-time RTI rollout. This
matters because it shows that the experimental zero-shot result is not
already latent in the pretrained checkpoint. RTI behavior must first be
learned through finetuning.

\begin{figure}[htb!]
    \centering
    \includegraphics[width=\linewidth]{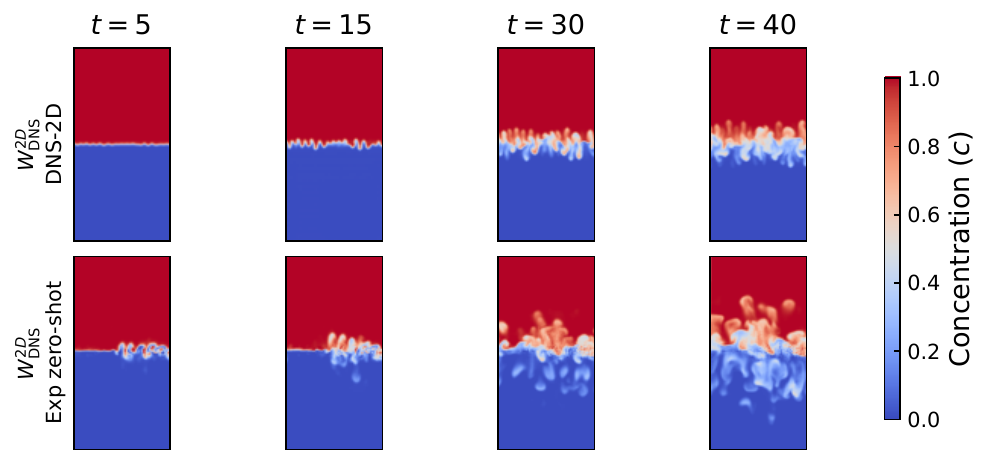}
    \caption{\textbf{Rollouts of the 2D DNS-specialized model $\WdnsTwoD$ on 2D DNS and experimental input.}
    Top row: rollout of $\WdnsTwoD$ initialized from 2D DNS input.
    Bottom row: zero-shot rollout of $\WdnsTwoD$ initialized from
    experimental input. Columns show representative times. After
    specialization to RTI on simulated 2D DNS slices, the model
    reproduces DNS-like growth in the top row, but develops markedly
    stronger mixing and interpenetration when initialized from
    experimental input in the bottom row. This comparison helps rule
    out the trivial alternative that any smooth propagator initialized
    from the experimental frames would drift toward the experimental
    late-time growth regime.}
    \label{fig:w2d_dns_zero_shot_dns2d_exp}
\end{figure}

Figure~\ref{fig:w2d_dns_zero_shot_dns2d_exp} shows the corresponding
rollouts of $\WdnsTwoD$. Once the model has been specialized on
simulated 2D RTI slices, its behavior changes qualitatively. When
initialized from 2D DNS input, $\WdnsTwoD$ produces the expected
DNS-like evolution of the interface. When initialized from experimental
input, it no longer behaves as a generic smooth propagator. Instead, it
develops substantially stronger growth and broader interpenetration than
in the DNS-like 2D case, in line with the upward shift in late-time
$\alpha$ reported in the main text. Taken together, these comparisons
show that the zero-shot experimental result depends on RTI
specialization. It is not a trivial consequence of architecture,
smoothness, or autoregressive propagation alone.

\clearpage

\subsection{Early experimental structure after second-stage finetuning}
\label{app:exp_ic}

\begin{figure*}[htb!]
    \centering
    \includegraphics[width=0.8\linewidth]{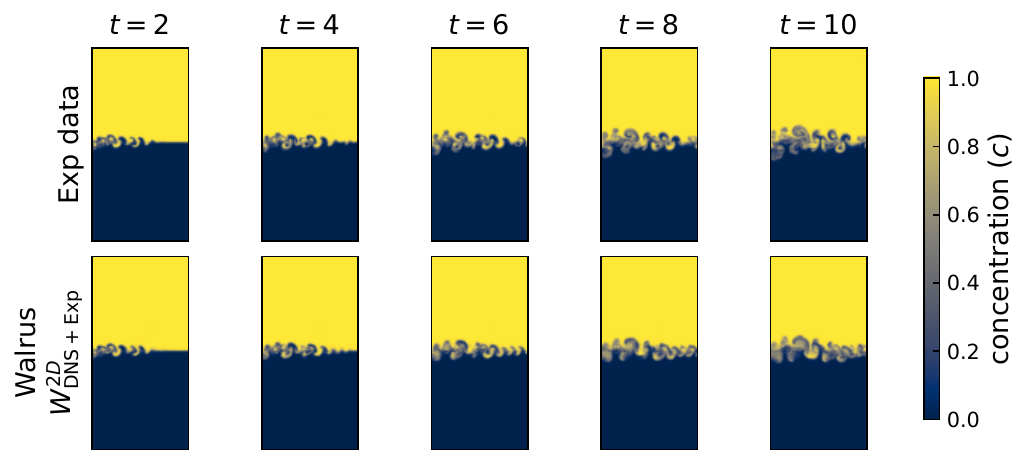}
    \caption{\textbf{Early experimental evolution and the effect of second-stage finetuning.}
    Top: held-out experimental concentration fields during the
    barrier-release transient and early instability growth. Bottom:
    corresponding rollouts of the experimentally adapted model
    $\Wexp$. Relative to the zero-shot results discussed in the main
    text, the second finetuning stage improves agreement with the
    release-driven early-time evolution while preserving the broader
    late-time growth regime.}
    \label{fig:exp_ic_rollout}
\end{figure*}

To clarify what the second experimental finetuning stage adds beyond
the zero-shot result, we compare early experimental frames with the
corresponding rollout of $\Wexp$ on held-out experimental data
(Fig.~\ref{fig:exp_ic_rollout}). The main improvement is not in the
late-time growth regime, which is already reached in zero shot, but in
the transient and intermediate stages of the evolution. In particular,
$\Wexp$ better follows the barrier-release-driven early-time structure
while preserving the larger late-time $\alpha$ values emphasized in
the main text.

\begin{figure}[htb!]
    \centering
    \includegraphics[width=0.6\linewidth]{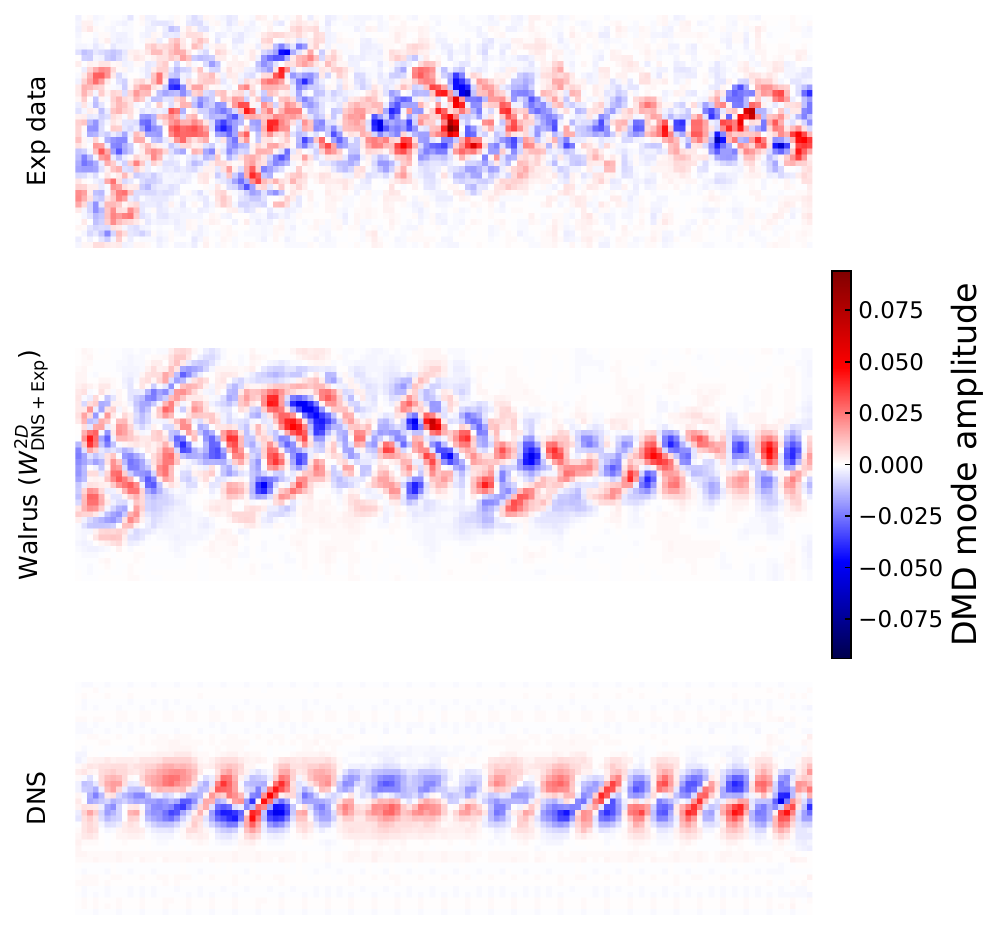}
    \caption{\textbf{Dominant early-time DMD mode in experiment, $\Wexp$, and the idealized DNS baseline.}
    The leading dynamic mode decomposition (DMD) mode of the early-time
    vorticity field highlights the large-scale anisotropic structure
    associated with experimental release. This structure is pronounced
    in the experimental data (top) but is not captured by the idealized
    DNS baseline (bottom), whose dominant mode remains more regular and
    columnar. After light experimental finetuning, $\Wexp$ (middle)
    shifts toward the experimental mode structure, supporting the view
    that a DNS-specialized RTI prior can be adjusted to
    experiment-specific release anisotropy.}
    \label{fig:early_time_dmd}
\end{figure}

A complementary view of the early experimental transient comes from
dynamic mode decomposition (DMD), which extracts coherent
spatiotemporal structures from a sequence of flow
fields~\cite{schmid2010dmd}. Given a snapshot matrix
\(X=[x_1,x_2,\dots,x_m]\), DMD approximates the evolution
\(x_{k+1}\approx A x_k\) and identifies spatial modes \(\phi_j\)
associated with characteristic temporal behavior. Here we apply DMD to
the early-time vorticity field in order to isolate the dominant
large-scale structure generated during the barrier-release transient.

Figure~\ref{fig:early_time_dmd} compares the leading early-time DMD
mode for the experimental data (top), the experimentally adapted model
$\Wexp$ (middle), and the idealized DNS baseline used for 2D RTI
specialization (bottom). The figure shows a cropped region around the interface only where the perturbations are the strongest. The experimental mode exhibits a broad,
anisotropic large-scale structure that is not naturally captured by the
idealized DNS, whose leading mode remains more regular and columnar.
After light experimental finetuning, $\Wexp$ shifts toward the
experimental mode structure. The dominant DMD mode is not intended as a
complete description of the flow; rather, it provides a compact summary
of the leading coherent early-time organization. In the present
context, that is precisely the relevant diagnostic, because the
barrier-release transient introduces anisotropic large-scale structure
that is absent from the idealized DNS used to obtain $\WdnsTwoD$.
\clearpage

\end{document}